\documentclass[twocolumn]{aa}  

\usepackage{graphicx}

\usepackage{txfonts}
\usepackage{xcolor}
\usepackage{comment}
\usepackage{array} 
\usepackage{amsmath}
\usepackage{amsfonts}
\usepackage{amssymb}
\usepackage[colorlinks=true,linkcolor=red,citecolor=blue,filecolor=magenta,urlcolor=blue]{hyperref}

\usepackage[normalem]{ulem}

\begin{document}

   \title{Identifying  compact symmetric objects with high-precision VLBI and \textit{Gaia} astrometry}
   \titlerunning{Identifying CSOs with \textit{Gaia} and VLBI}

   \author{T. An \inst{1,2}\fnmsep\thanks{Corresponding author: antao@shao.ac.cn},
          Y. Zhang \inst{1,2},
          S. Frey\inst{3,4,5},
          W.A. Baan\inst{6,1,7} \and
          A. Wang\inst{1}}

   \institute{Shanghai Astronomical Observatory, Key Laboratory of Radio Astronomy, CAS, 80 Nandan Road, Shanghai 200030, China
        \and
        State Key Laboratory of Radio Astronomy and Technology, A20 Datun Road, Chaoyang District, Beijing, P. R. China
         \and
             Konkoly Observatory, HUN-REN Research Center for Astronomy and Earth Sciences, Konkoly Thege Mikl\'{o}s \'{u}t 15-17, H-1121 Budapest, Hungary 
        \and
             CSFK, MTA Centre of Excellence, Konkoly Thege Mikl\'os \'ut 15-17, H-1121 Budapest, Hungary
        \and 
            Institute of Physics and Astronomy, ELTE E\"{o}tv\"{o}s Lor\'{a}nd University, P\'{a}zm\'{a}ny P\'{e}ter s\'et\'{a}ny 1/A, H-1117 Budapest, Hungary
        \and 
            Xinjiang Astronomical Observatory, CAS, 150 Science-1 Street, \"Ur\"umqi, Xinjiang 830011, P.R. China
        \and
            Netherlands Institute for Radio Astronomy, ASTRON, Oude Hoogeveensedijk 4, 7991PD Dwingeloo, The Netherlands
         }

   \date{Received 2025; accepted 2025}

  \abstract
   {
   Compact symmetric objects (CSOs) trace the earliest phases of radio-galaxy growth; however, robust classification is difficult when radio cores are weak or invisible. 
   }
   {We aim to develop and test a \textit{Gaia}+VLBI approach that utilizes the high-precision optical astrometry of  \textit{Gaia} together with the high-resolution imaging of very long baseline interferometry (VLBI) to reliably locate the central engine and classify CSOs.} 
   {We analysed 40 literature CSO candidates by overlaying \textit{Gaia} Data Release 3 (DR3) positions on VLBI maps and by examining spectral index distributions, whole-source variability, and hotspot kinematics over up to 25 years. A source is classified as a CSO when the \textit{Gaia} centroid lies between two steep-spectrum lobes; alignment with one end of the radio structure indicates a core--jet source. }
   {Our method yields 20 confirmed CSOs, ten core--jet sources, and ten ambiguous cases affected by significant optical--radio positional offsets or limited data. The confirmed CSOs show low integrated variability, hotspot advance speeds typically $<0.5 \, c$, where $c$ denotes the speed of light (with a few mildly relativistic cases), and kinematic ages of $\approx 20-2000$~yr. Five nearby CSOs show systematic \textit{Gaia}--VLBI offsets despite the CSO-like morphology, likely reflecting host-galaxy environments and \textit{Gaia} astrometric systematics. We find a clear dichotomy with radio power: high-power CSOs ($P_{1.4\,\mathrm{GHz}} > 10^{26.5}\,\mathrm{W}\,\mathrm{Hz}^{-1}$) tend to be larger and host faster hotspots, while many low-power systems remain sub‑kiloparsec and environmentally confined.  }
   {\textit{Gaia}+VLBI registration is a powerful method for CSO classification,  especially where radio cores are faint. 
   The observed power-size-velocity-age relations support distinct multiple evolutionary tracks, with high-power CSOs plausibly growing into large radio galaxies, while low-power CSOs appear confined by their host galaxy environments. Taken together, our results indicate that CSO evolution is shaped not only by intrinsic jet power, but also by host--galaxy environment and the duty cycle of the central engine.
   High-sensitivity observations of low-power CSOs will be crucial to map the full diversity of formation channels and evolutionary pathways of radio galaxies.
   }


   \maketitle

\section{Introduction} \label{sec:intro}

Compact symmetric objects (CSOs) represent a key early stage in radio galaxy growth, characterized by mirror-symmetric double-lobed morphology within a projected size of  $\lesssim 1 $ kiloparsec (kpc) \citep{1980ApJ...236...89P, 1982A&A...106...21P,1981ApJ...248...61P, 1988ApJ...328..114P}. These sub-kiloparsec, symmetric radio sources provide clean laboratories for jet--interstellar medium (ISM) coupling and nascent active galactic nucleus (AGN) feedback with minimal projection effects \citep{2016A&ARv..24...10T, 2018MNRAS.475.3493B, 2021A&ARv..29....3O}.

The physical nature of CSOs has been framed by three primary scenarios, which are as follows. 
(I) The young radio galaxy scenario suggests CSOs represent the earliest evolutionary stage of radio galaxies \citep[e.g.,][]{1980ApJ...236...89P, 1996ApJ...460..634R,1998A&A...337...69O}, supported by the observed ``luminosity--size'' relationship \citep{1982IAUS...97...21B}, which indicates progression from CSOs to larger scale radio sources \citep[e.g.,][]{1995A&A...302..317F,2000MNRAS.319..445S,2010MNRAS.408.2261K,2012ApJ...760...77A}. However, this evolutionary path may only account for a small fraction of the overall CSO population, primarily those with high-power jets ($L_{1.4\,\mathrm{GHz}} > 10^{26.5}\,\mathrm{W\,Hz^{-1}}$), while the fate of those with lower power and unstable jets is uncertain.
(II) In the frustration model,  CSOs are confined by dense ISM in their host galaxies \citep[e.g.,][]{1984AJ.....89....5V, 1984Natur.308..619W, 1991ApJ...380...66O, 1993ApJ...402...95D}, especially at low radio power, potentially linking them to Fanaroff--Riley (FR) type 0 (FR0) radio galaxies \citep{1974MNRAS.167P..31F, 2018A&A...609A...1B, 2019MNRAS.482.2294B}. Even lower power radio sources \citep{2024MNRAS.529.1472C} suggest a potential link to a previously unexplored low-power CSO population, where jet--ISM interactions significantly shape radio morphology.
(III) The transient-phenomenon model suggests some CSOs may be short-lived, resulting from intermittent or episodic AGN activity. It is important to distinguish between ``intermittent'' (e.g., repeating cycles caused by accretion-disk instabilities, \citealt{2002ApJ...576..908J, 2009ApJ...698..840C, 2010ASPC..427..326S, 2009ApJ...701L..95W}) and ``episodic'' events (one-time phenomena such as tidal disruption events, TDEs, \citealt{1996ApJ...460..612R, 2024ApJ...961..242R}). While the episodic TDE model can explain certain short-lived sources, it struggles to account for CSOs that evolve into larger radio galaxies or show multiple activity episodes \citep[e.g.,][]{2000MNRAS.315..395S, 2000MNRAS.315..381K, 2009A&A...506L..33M, 2011MNRAS.414.1397J, 2019A&A...622A..13M, 2023MNRAS.522.3877O}. The presence of double-double radio lobes or remnant emission suggests that CSOs may undergo recurrent activity cycles \citep{2010A&A...510A..84M, 2013MNRAS.436.1595K}, requiring models that explain both transient and long-term evolutionary processes.

To understand radio-galaxy formation, precise quantification of CSO dynamics, ages, and triggering rates is crucial to establish their relationship with large-scale radio galaxies. CSOs provide unique insights into the birth and growth of radio galaxies through several key aspects: their compact size and youth provide direct windows into early-stage jet activity and jet--ISM interactions, while their symmetric double-lobed morphology, oriented close to the sky plane, minimizes relativistic beaming effects and reveals intrinsic jet properties.

Compact symmetric objects are essential for understanding AGN feedback mechanisms and galaxy evolution. They serve as laboratories for studying three critical processes: initial jet--ISM coupling in galactic nuclei, jet physics with minimal projection effects, and diverse evolutionary pathways spanning from high-power sources evolving into large radio galaxies to confined low-power sources. These characteristics make CSOs crucial to understanding how AGN feedback operates and influences galaxy evolution across cosmic time.

Despite their importance, identifying and understanding the physical nature of CSOs remains challenging. The primary difficulty lies in detecting their often weak or obscured radio cores \citep{2005ApJ...622..136G}, caused by several factors. First, minimal relativistic beaming due to large inclination angles reduces core emission \citep{1999ApJ...521..103P,2006A&A...450..959O,2001ApJ...550..160M,2008A&A...487..885O,2012ApJS..198....5A,2013A&A...550A.113W}. Second, dense ionized gas in nuclear regions can cause significant attenuation through synchrotron self-absorption or free-free absorption \citep{1998PASP..110..493O,2009AN....330..120F}. Third, frequency-dependent opacity effects can shift apparent core positions, complicating identification.

Traditional CSO identification has relied on morphological and spectral criteria derived from high-resolution, very-long-baseline-interferometry (VLBI) observations. Key indicators include a symmetric double-lobed structure, steep-spectrum lobes, and low variability \citep[e.g.,][]{1996ApJ...463...95T,2000ApJ...534...90P,2007ApJS..171...61H}. However, distinguishing genuine CSOs from core--jet sources exhibiting similar structures remains difficult in the absence of clear core detection. While  multifrequency VLBI observations provide supplementary evidence, uncertainties in CSO variability, dynamics and life cycles limit our ability to construct unbiased CSO samples.

To address these challenges, we introduce a new method combining high-precision optical astrometry from the \textit{Gaia} space mission \citep{2016A&A...595A...1G} with high-resolution VLBI imaging to reliably identify genuine CSOs. \textit{Gaia} Data Release 3 \citep[DR3,][]{2023A&A...674A...1G} provides unprecedented astrometric precision ($\sim$0.04 mas for bright sources with apparent optical magnitude $G<14^{\mathrm{mag}}$ and $\sim 0.7$ mas at $G=20^{\mathrm{mag}}$) for approximately 1.6 million quasar-like objects. When overlaid on VLBI maps, these precise optical positions pinpoint AGN central engines, even when radio cores are weak or undetectable.

This study applies this method to 40 CSO candidates selected from existing VLBI surveys (Section~\ref{sec:method}). 
By overlaying \textit{Gaia} DR3 positions on multi-epoch, multifrequency VLBI maps, we were able to classify these objects and study their properties in  detail.

The structure of this paper is as follows. Section~\ref{sec:method} presents the methodology and parent sample. Section~\ref{sec:result} presents the identification of CSOs and core--jet sources using the \textit{Gaia}+VLBI method. Section~\ref{sec:disc} discusses the radio properties of the confirmed CSOs. 
The results are summarised in Section~\ref{sec:summary}. Throughout this paper, we assume a standard flat $\Lambda$ cold dark matter cosmology with $H_{\rm{0}} = 70$~km\,s$^{-1}$\,Mpc$^{-1}$,  $\Omega_{\rm{m}} = 0.3$, and $\Omega_{\Lambda} = 0.7$. The radio spectrum is defined as $S_\nu \propto \nu^{-\alpha}$, where $S_\nu$ is the flux density at the observing frequency, $\nu$; and $\alpha$ is the spectral index.

\section{Method and samples} 
\label{sec:method}

\subsection{CSO identification criteria}

We used four diagnostics for CSO identification:

\begin{enumerate}
    \item Morphology.  A compact, mirror-symmetric, double-lobed radio structure with projected linear size of $<1$~kpc and a lobe-brightness ratio of $<$10:1; 

    \item Spectral properties. Steep-spectrum lobes and a flat or inverted-spectrum core (when detected); convex or overall steep radio spectra are common;
   
    \item Low integrated variability. Whole-source variability is typically $<20\%$;  
    
    \item Hotspot kinematics. Sub-relativistic hotspot advance speeds, typically $<0.5\, c$, where $c$ denotes the speed of light. 
\end{enumerate}

In principle, a confirmed CSO satisfies all four criteria. In practice, items (3) and (4) require homogeneous, long-term monitoring and are not always available. We therefore used (1) and (2) for the initial classification and treated (3) and (4) as confirming tests where data exist. If any diagnostic conflicted with CSO expectations, we adopted the conservative label of CSO candidate (CSOc). Table \ref{tab:cso_diag_summary} summarises, for each source, the diagnostics available and their outcomes, including the whole-source variability measurement.

\setlength{\tabcolsep}{2pt}
\begin{table}
\caption{Diagnostics applied to CSO and CSOc classes.}
\label{tab:cso_diag_summary}
\centering
\small
\begin{tabular}{lccccc}
\hline\hline
Name & Cl. & \textit{Gaia}+VLBI\tablefootmark{a} & M.+S.\tablefootmark{b} & $V_{2.3}$\tablefootmark{c} & $v_{\rm HS}$\tablefootmark{d}  \\
(1) & (2) & (3) & (4) & (5) & (6) \\
\hline
J0048+3157 & CSO  & $\times$     & $\checkmark$ & / & / \\
J0741+2706 & CSO  & $\checkmark$ & $\checkmark$ & / & /  \\
J0832+1832 & CSO  & $\checkmark$ & $\checkmark$ & 0.044 ($\checkmark$) & $\checkmark$  \\
J1110+4817 & CSO  & $\checkmark$ & $\checkmark$ & 0.042 ($\checkmark$) & $\checkmark$ \tablefootmark{e}  \\
J1111+1955 & CSO  & $\checkmark$ & $\checkmark$ & 0.036 ($\checkmark$) & $\checkmark$  \\
J1158+2450 & CSO  & $\checkmark$ & $\checkmark$ & 0.046 ($\checkmark$) & $\checkmark$ \\
J1234+4753 & CSO  & $\checkmark$ & $\checkmark$ & 0.036 ($\checkmark$) & $\checkmark$ \\
J1244+4048 & CSO  & $\checkmark$ & $\checkmark$ & 0.024 ($\checkmark$) & $\checkmark$\tablefootmark{e} \\
J1247+6723 & CSO  & $\times$     & $\checkmark$ & 0.032 ($\checkmark$) & $\checkmark$  \\
J1256+5652 & CSO  & $\checkmark$ & $\checkmark$ & 0.062 ($\checkmark$) & / \\
J1310+3403 & CSO  & $\checkmark$ & $\checkmark$ & /                    & / \\
J1326+3154 & CSO  & $\checkmark$ & $\checkmark$ & /                    & / \\
J1358+4737 & CSO  & $\checkmark$ & $\checkmark$ & 0.002 ($\checkmark$) & $\checkmark$ \\
J1407+2827 & CSO  & $\times$     & $\checkmark$ & 0.179 ($\checkmark$) & $\checkmark$ \\
J1511+0518 & CSO  & $\times$     & $\checkmark$ & 0.077 ($\checkmark$) & $\checkmark$ \\
J1602+2418 & CSO  & $\checkmark$ & $\checkmark$ & 0.085 ($\checkmark$) & $\times$\tablefootmark{e} \\
J1755+6236 & CSO  & $\checkmark$ & $\checkmark$ & / & /  \\
J1815+6127 & CSO  & $\checkmark$ & $\checkmark$ & 0.093 ($\checkmark$) & /\tablefootmark{e}  \\
J1945+7055 & CSO  & $\times$     & $\checkmark$ & 0.109 ($\checkmark$) & $\checkmark$ \\
J2355+4950 & CSO  & $\checkmark$ & $\checkmark$ & / & /  \\
\hline
J0119+3210 & CSOc & $\times$     & $\checkmark$ & / &  $\checkmark$\tablefootmark{g}  \\
J0650+6001 & CSOc & $\times$     & $\checkmark$ & / & $\checkmark$ \\
J0831+4608 & CSOc & $\times$     & $\checkmark$ & / & /  \\
J0906+4636 & CSOc & $\times$     & $\checkmark$ & / & / \\
J0943+1702 & CSOc & $\checkmark$ & $\checkmark$ & / & $\times$  \\
J1148+5924 & CSOc & $\times$     & $\checkmark$ & / & / \\
J1254+1856 & CSOc & $\times$     & $\checkmark$ & / & / \\
J1309+4047 & CSOc & $\times$     & $\checkmark$ & / & / \\
J1559+5924 & CSOc & $\times$     & $\checkmark$ & / & / \\
J1823+7938 & CSOc & $\times$     & $\checkmark$ & / & $\checkmark$ \\
\hline
\end{tabular}
\tablefoot{
Columns (1)--(2) are source name, classification, and columns; (3)--(6) are diagnostics. 
\tablefoottext{a}{$\checkmark$ = \textit{Gaia} centroid near the midpoint between two lobes; $\times$ = ``large optical--radio offset.''} 
\tablefoottext{b}{$\checkmark$ is applied when a two-sided, edge-brightened structure with steep-spectrum lobes and flat or inverted core is established in VLBI maps.} 
\tablefoottext{c}{Whole-source variability $V_{\rm 2.3}=(S_{\max}-S_{\min})/(S_{\max}+S_{\min})$ at 2.3 GHz using matched ($u, v$) coverage; ``$\checkmark$'' if $V_{\rm 2.3}<0.20$.} 
\tablefoottext{d}{ ``$\checkmark$'' if hotspot separation speed is $v_{\rm HS}<1.0\,c$; ``$\times$'' if $v_{\rm HS}\ge 1.0\,c$; ``/'' if not measured. } 
\tablefoottext{e}{Values are limits or pattern-/orientation-affected; see kinematics discussion in Section \ref{sec:disc}.} \\
}
\end{table}

\subsection{VLBI CSO sample} \label{sec:sample}

\begin{figure*}
\centering
\includegraphics[width=0.9\linewidth]{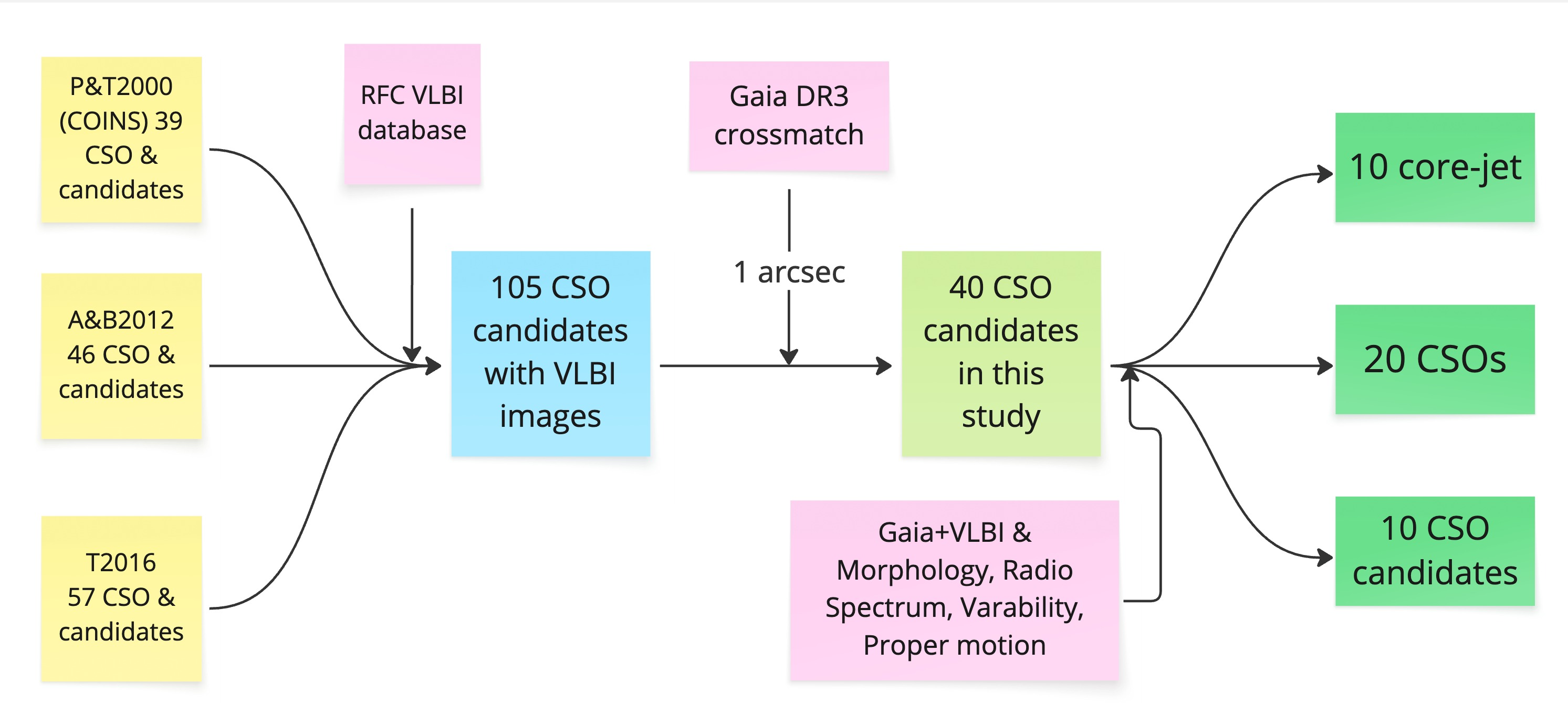}   
\caption{Flowchart of our sample selection procedure. P\&T2000: \citet{2000ApJ...534...90P}; A\&B2012: \citet{2012ApJ...760...77A}; T2016: \citet{2016MNRAS.459..820T}. The Radio Fundamental catalogue (RFC):  \url{http://astrogeo.org/rfc/} \citep{2024arXiv241011794P}. \textit{Gaia} DR3: \url{https://www.cosmos.esa.int/web/gaia/data-release-3}. }
\label{fig:sketch}
\end{figure*}

Our CSO candidate sample combines data from three representative CSO studies: the COINS (CSOs observed in the northern sky) sample \citet{2000ApJ...534...90P}, the A\&B2012 sample \citet{2012ApJ...760...77A}, and the T2016 sample \citet{2016MNRAS.459..820T}. These studies yielded
over 140 candidates based on symmetric morphology and steep or convex radio spectra.
After removing redundant sources (Fig.~\ref{fig:sketch}), we established a consolidated sample of 105 CSO candidates. While not comprehensive in sky coverage or flux-density limits, this dataset represents the most credible collection of CSOs and candidates confirmed through multiple criteria (primarily morphology and radio spectra), providing an effective testbed for our CSO identification technique. The sample spans a broad range of redshifts (0.015 to 1.791, median 0.269; Fig. S1 in Supplementary Material\footnote{Additional figures, tables, and methods are available at Zenodo (\url{https://doi.org/10.5281/zenodo.17278723}). Figure and table numbering in Supplementary Material follows the S-scheme (e.g., Fig. S1, Table S1)}).

We first selected 40 CSO candidates using \textit{Gaia}+VLBI registration (Fig.~\ref{fig:sketch}), restricted to sources that already satisfy the morphological and radio spectral criteria. For objects with variability and/or hotspot motion measurements, any robust inconsistency with these diagnostics leads to a conservative downgrade to a CSOc.
Some sources show large \textit{Gaia}--VLBI offsets; however, strong independent-literature evidence supports their classification as CSOs (Table \ref{tab:cso_diag_summary}). Conversely, some objects show good \textit{Gaia}--VLBI alignment, yet fail the low-variability or sub-relativistic hotspot speed expectations for standard CSOs, and are therefore tentatively listed as CSOc. We note minor discrepancies with classifications in Readhead et~al. (in prep.; private communication) for a few objects; these do not affect our general conclusions and will be revisited as community lists converge.

\subsection{\textit{Gaia}+VLBI sample} \label{sec:Gaia+VLBI}

We introduce a novel methodology that combines high-precision optical astrometry from \textit{Gaia} with high-resolution VLBI imaging to reliably identify CSO cores. Previous comparisons between \textit{Gaia} and VLBI astrometry have focused mainly on radio--optical positional offsets in core--jet systems \citep{2017MNRAS.467L..71P}. More recently, exploratory studies have begun to test the joint \textit{Gaia}+VLBI approach for individual GHz-peaked-spectrum quasars \citep[][]{2024MNRAS.528.1697K}. Our work constitutes the first systematic application of this technique to a large CSO sample, directly addressing the long-standing difficulty of CSO core identification.

We cross-matched 105 VLBI-selected CSO candidates with \textit{Gaia} DR3 sources using a conservative $1''$ search radius and identified 40 robust optical counterparts (Table \ref{tab:sample}). The \textit{Gaia} optical centroid probes the immediate vicinity of the accretion disk and the central supermassive black hole (SMBH), thereby providing a stable reference for the AGN core position (see Sect. S2). When registered against VLBI maps with sub-milliarcsecond (mas) resolution, these optical centroids enable precise localization of the central engine, even when the radio core is faint or absorbed. The modest match fraction (38\%) further indicates that many CSO nuclei are intrinsically optically faint or heavily obscured.

The cross-matched subset comprises a heterogeneous mix of quasars ($\sim 45\%$) and galaxies ($\sim 55\%$). From the redshift distribution (Fig. S1), most CSOs and candidates cluster at low and modrate redshifts ($z<1.0$), which is largely set by the \textit{Gaia} detection limit: galaxies beyond $z \sim 1$ are often too faint for  reliable \textit{Gaia} astrometry. In contrast, core--jet sources show a broader redshift range.  

We performed a detailed analysis using the latest DESI Data Release 1 \citep{2025arXiv250314745D} to assess the detection rates of CSO candidates. Out of our parent sample (105 sources), 63 fall within the DESI footprint, while 42 do not. Among the sources within the DESI footprint, those without \textit{Gaia} counterparts (35 sources) show a low detection rate ($\sim 20\%$) in DESI, with detected sources spanning a redshift range of $z =$ 0.2--1.5. In contrast, \textit{Gaia}-detected sources exhibit a higher detection rate ($\sim 50\%$) in DESI. This stark difference in detection rates supports the notion that host-galaxy detection is significantly affected by distance. The \textit{Gaia} non-detections recovered by DESI at higher redshifts confirm that many of these optically faint sources are distant galaxies, not intrinsically faint nearby objects. This result highlights the optical-selection bias and its effect on our CSO sample.

To assess astrometric quality, we screen \textit{Gaia} solutions using $G$ magnitude, the astrometric excess noise (AEN), and its significance (AENS) (Table S2), and we treat significant \textit{Gaia}--VLBI offsets with caution. Particularly for faint sources ($G\gtrsim20$ mag) or extended/structured hosts, low signal-to-noise ratios, asymmetry, or blending can bias the optical centroid and produce spurious \textit{Gaia}--VLBI offsets, analogous cross-identification uncertainties were reported by \citet{2019ApJ...871..143P}. In such cases, we require independent radio diagnostics before rejecting a CSO classification. This behavior is exemplified by several nearby, well-studied CSOs in Table \ref{tab:cso_diag_summary} (e.g., J1407+2827/OQ~208, J1511+0518, J1945+7055), which have strong literature CSO identifications yet show large \textit{Gaia}--VLBI offsets attributable to \textit{Gaia} astrometric systematics rather than genuine optical--radio displacements (see also Section \ref{sec:cso_confirm}).

This combined approach offers several advantages over traditional, radio-only identification: (i) precise nuclear localization even when the radio core is weak or obscured; (ii) reliable discrimination between CSOs and core--jet sources; (iii) an independent verification of the orientation of the radio structure; and (iv) improved kinematic measurements by anchoring component motions to a fixed, physically motivated reference point.

\section{Results} \label{sec:result}

Our application of the \textit{Gaia}+VLBI method yielded classifications for 30 of 40 CSO candidates, demonstrating the effectiveness of combining high--precision optical astrometry with radio imaging. We confirmed 20 sources as bona fide CSOs, and identified ten as core--jet AGN (Figures S2--S3, Tables S3--4). The remaining ten sources (Fig. S4, Table S5) are ambiguous due to significant radio--optical positional discrepancies.

\subsection{Confirmed CSOs}
\label{sec:cso_confirm}

Among the 20 confirmed CSOs, 15 show direct \textit{Gaia}+VLBI alignment; the remaining five are well studied in the literature and are confirmed as CSOs by independent evidence, despite larger \textit{Gaia}--VLBI offsets.. Most exhibit classic double-lobed structures (e.g., J1111+1955, J1326+3154), while several show complex morphologies indicating jet-ISM interactions. Sources such as J0832+1832, J1158+2450, and J1815+6127 display asymmetric brightness where components closest to the core appear brightest, suggesting strong environmental influence. Additional evidence of the environment includes jet deflection (J0832+1832, J1815+6127), bending (J1158+2450), and clumpy morphology (J1244+4048). Multifrequency coverage is essential: 2.3-GHz data reveal extended, steep-spectrum features in J0741+2706, J0832+1832, J1256+5652, and J2355+4950 that are too faint at higher frequencies owing to spectral steepening and Doppler deboosting.

Lobes generally have steep spectra ($\alpha > 0.5$) from aging relativistic electrons. Central regions generally show flatter spectra, steepening toward the lobes, consistent with opacity and/or ongoing particle injection near the core, followed by spectral aging as the plasma flows outward. Hotspots in young radio sources are expected to be optically thin with steep spectra ($\alpha \gtrsim 0.5$). Flat or inverted spectral indices in some regions (e.g., NE jet of J1158+2450) are best explained by opacity effects (synchrotron or free-free absorption) or contamination from unresolved core/inner jet emission \citep{1996ApJS..107...37T,2014MNRAS.438..463O}.

Kinematic analysis (Table~\ref{tab:pm_fit}) reveals diverse hotspot separation speeds in our CSO sample. Most sources display sub-relativistic motions ($< 0.5\,c$), which are characteristic of symmetric jet expansion in young radio galaxies. J1602+2418 shows mildly relativistic speed of $(v_{\rm HS} = 2.54 \pm 0.45)\,c $, though with substantial uncertainties. J1111+1955 may exhibit an unusually high apparent speed that requires confirmation with longer time baseline. Estimated kinematic ages span $\sim 80$ years (J1602+2418) to over 1600 years (J1234+4753), revealing diverse evolutionary stages. Three sources (J1110+4817, J1256+5652, J1815+6127) show apparent negative speeds, likely from structure changes or projection effects \citep{2012ApJS..198....5A}.

Five nearby ($z \leq 0.1$) CSOs --- J0048+3157 (NGC~262, Mrk~348), J1247+6723, J1407+2827, J1511+0518, and J1945+7055 --- show large radio--optical offsets despite characteristic CSO morphology and spectra. In J1407+2827 (OQ 208, \citealt{1997A&A...318..376S, 2012ApJS..198....5A, 2013A&A...550A.113W}),  the \textit{Gaia} position coincides with a radio lobe, likely due to bright optical emission from a jet--ISM interaction site. 
Host-galaxy factors (central asymmetries, host spectral components, and dust extinction) can shift optical centroids relative to VLBI positions \citep{2019ApJ...871..143P}, and starburst/merging hosts can further increase \textit{Gaia} uncertainties. These cases illustrate limitations of the \textit{Gaia}+VLBI method for nearby, optically weak or obscured nuclei, but their CSO classification remains secure from morphology and spectra.

The projected linear sizes of the confirmed CSOs range from 3.6 to 290~pc, all within the canonical $<1$~kpc  CSO scale (Table~\ref{tab:ab2012fig}). This diverse distribution enables tests of CSO evolution from initial jet launch to early growth of radio structure. 

\subsection{Core--jet sources}

We identified ten core--jet sources in which \textit{Gaia} positions coincide with the brightest terminal component of an elongated radio structure, which is consistent with relativistic beaming (Fig. S3). 
A frequency-dependent core shift (systematic position changes with observing frequency due to synchrotron self-absorption gradients along the jet) modulates these alignments \citep{1999ApJ...521..509L}. Higher radio frequencies probe regions closer to the SMBH, while the optical centroid typically lies even farther upstream \citep{2008A&A...483..759K}. In several sources (e.g., J0753+4231, J1311+1417), the optical position  aligns with a weaker upstream component rather than the brightest radio feature, marking the true core locations.
In J0003+2129, the \textit{Gaia} position is displaced toward the counter-jet relative to the radio peak; the absence of radio emission at the optical position is consistent with strong opacity at the jet base, whereas the optical traces the unobscured nucleus.
Most optical-radio offsets are $<1$ mas and compatible with astrometric uncertainties. The $\sim13$ mas offset for J2022+6136 likely reflects its faintness and large astrometric uncertainty ($G = 20.68$). 
Kinematic analysis  (Table S6) shows apparent jet speeds of $0.68\,c$--$5.43\,c$, well above the subrelativistic CSO hotspot motions, confirming their relativistic jets.

\subsection{CSO candidates}

Ten sources show \textit{Gaia}--VLBI offsets larger than the combined astrometric uncertainties, implying that the \textit{Gaia} detections may trace unrelated foreground/background objects or extended host features rather than the AGN cores \citep{2019MNRAS.490.5615B}.
J0831+4608 and J1254+1856 were observed at only one epoch and frequency; their morphologies are CSO-like, but confirmation requires multi-epoch, multifrequency VLBI to measure spectra and proper motions. J0943+1702 shows a symmetric double at 2.3~GHz with the \textit{Gaia} position near the midpoint. At 8.7~GHz the structure resolves into N--C--S, with C having flat-spectrum and N/S steep-spectrum. Total-flux variability appears modest (S band), but the apparent $\sim2\,c$ hotspot speed is atypical of CSOs and may reflect projection or pattern speeds; we therefore retain a conservative CSO-candidate label. J1823+7938 displays a double-lobe morphology at 2.3~GHz; at 8.4~GHz, the \textit{Gaia} centroid lies near the E component, which shows flat spectrum, favouring a core+jet geometry. Its roughly symmetric morphology, slow (subluminal) expansion, and low S-band variability are all consistent with a CSO, but the \textit{Gaia} offset from the geometric midpoint motivates a conservative CSO-candidate classification.

J1559+5924 shows a bright central component with two diffuse extensions, resembling a mini FR~I; it remains a CSO candidate, possibly a low-power, two-sided-jet system.
J0119+3210, J0906+4636, and J1148+5924 display complex emission without compact, edge-brightened lobes. Their component brightness temperatures ($10^8$--$10^9$~K) and radio powers ($L_{1.4\,\mathrm{GHz}} \sim 10^{22}$--$10^{24}$~W~Hz$^{-1}$) exceed thermal star-formation values yet fall below typical blazar cores, consistent with non-thermal, low-power jets rather than classical CSOs. Also 
J0650+6001 remains ambiguous: its \textit{Gaia} position near the southern component suggests a core--jet source, yet the overall GHz-peaked spectrum and slow jet motion ($0.35\,c$) are CSO-like \citep{2010MNRAS.406..529O, 2012ApJS..198....5A, 2019ApJ...874...43L}; its strong total-flux variability (from $\sim$1000~mJy in 2012 to $\sim$200~mJy in 2021) argues against a standard CSO.
These cases highlight classification limitations and the need for a multi-criterion approach. Targeted, multi-epoch and multiband VLBI will be decisive for confirming or rejecting CSO status.

\section{Discussion} \label{sec:disc}

\subsection{Efficacy and validation of the \textit{Gaia}+VLBI method}

Combining \textit{Gaia} astrometry with VLBI imaging enables robust core identification in compact radio sources. Our confirmation of 20 CSOs demonstrates this method's effectiveness, providing core localization even when radio cores are faint or obscured, clear separation of CSOs from core--jet systems, independent checks on structure orientation, and improved kinematic baselines.

This method complements radio-only catalogues such as the recent bona fide CSO sample \citep{2024ApJ...961..240K, 2024ApJ...961..241K}: while that catalogue of 79 CSOs, including 43 edge-brightened CSOs (CSO 2 class) with redshifts, applies strict radio-based criteria, our optical--radio registration provides an additional diagnostic. The $\sim 60\%$ overlap (12/19) between our confirmed CSOs and their list supports the approach. Two cases that are not consistent with the other classifications are classified as core-jet sources (J1816+3457 and J2022+6136), which highlights the known ambiguities in CSO identification.
Several important limitations must be considered:
\begin{itemize}
    \item Source confusion from foreground or background objects and extended emission from host galaxies can affect centroid positions.
    \item Effectiveness decreases for nearby galaxies with complex optical structure.
    \item Bias against heavily dust-obscured or optically faint nuclei.
    \item Inherent limitation to optically bright sources, restricting application to radio--quiet or very compact sources, particularly at high redshifts.
\end{itemize}

To mitigate these limitations, we recommend a multiwavelength approach combining data across different bands, careful assessment of host galaxy properties, and integration with traditional radio--based classification methods. Despite caveats, \textit{Gaia}+VLBI advances early-stage CSO identification where radio-only methods can be inconclusive. Future work will scale to a $>20,000$-source cross-match with Astrogeo VLBI, enabling quantitative tests of selection effects, environments, and CSO evolution across diverse populations.

\subsection{Radio morphology and spectral properties}\label{sec:radio-morph-spec}

Our high-resolution VLBI observations combined with \textit{Gaia} astrometry reveal diverse morphological and spectral properties that provide key insights into CSO physics and evolution.

The majority ($\sim70\%$) of confirmed CSOs exhibit a classical double-lobed structure with edge-brightened hotspots (e.g., J1111+1955, J1358+4737, J1602+2418), supporting theoretical models where young jets propagate symmetrically through a dense ISM. The emission typically becomes detectable when the jet terminates in a shock, efficiently converting kinetic energy into relativistic particles  \citep{1974MNRAS.169..395B, 1984RvMP...56..255B, 1995MNRAS.277..331S, 1989A&A...219...63M}. 
For high-power ($P_{\rm 1.4GHz} > 10^{26.5}$ W~Hz$^{-1}$) jets, kinetic energy is rarely lost as radiation during jet propagation, which often makes the core and inner jet invisible. 

Only six CSOs (J0741+2706, J1158+2450, J1234+4753, J1244+4048, J1256+5652, and J1310+3403) are detected with central radio components, which likely represent unresolved inner jet regions rather than cores. 
Several CSOs display complex morphologies indicating strong jet--ISM interactions. For example, J0832+1832 and J1815+6127 show multiple hotspots suggesting jet deflection or impact with dense ISM clouds \citep{2002MNRAS.331..323M}, or jet head impacting multiple sites at the interface \citep{2002ApJS..141..337C, 2012ApJS..198....5A}. 

The dramatically bent jet in J1158+2450 provides direct evidence of how strong jet--ISM interactions can sculpt the CSO morphology \citep{2016MNRAS.461..967M}. 
Similarly, the elongated, clumpy jet structure in J1244+4048 likely results from jet instabilities induced by entrainment of the ISM \citep[e.g.,][]{1984ApJ...286...68B}. 
In J1256+5652 (Mrk 231), we observe an east-west oriented jet within $2$~mas of the core, but extended emission at $20-30$~mas scales \citep{2021MNRAS.504.3823W}, 
hinting at multiple episodes of AGN activity and challenging simple, monotonic evolution models \citep{2009BASI...37...63S, 2011MNRAS.410..484B,2020MNRAS.499.1340O,2007MNRAS.382.1019B}. 
Such recurrent activity (Section \ref{sec4.5}), observed in numerous compact radio sources, reveals key aspects of both AGN feedback processes and long-term evolution of radio galaxies \citep{2000MNRAS.315..395S, 2012MNRAS.424.1061K, 2019MNRAS.486.5158N}.

The spectral properties provide additional information on the physical conditions in these sources. The overall radio spectra typically show a convex shape with a turnover near $5$~GHz, attributed to synchrotron self-absorption in compact, high-density regions \citep{1998PASP..110..493O}. This characteristic is observed in both low-redshift ($z<1$) and high-redshift ($z\geq 1$) CSOs \citep{2021MNRAS.508.2798S}.

The central regions generally exhibit flatter spectra, steepening toward the lobes (e.g., J1234+4753 and J1244+4048); this is consistent with theoretical models of continuous particle acceleration in central regions and is followed by spectral aging as the plasma flows outward \citep{1970ranp.book.....P}. The observed morphological and spectral diversity indicates that CSOs are not merely scaled-down versions of larger radio galaxies \citep{2012ApJ...760...77A}, they represent crucial laboratories for studying jet-ISM interactions, shock physics, and AGN feedback in action \citep{2016A&ARv..24...10T}.

\subsubsection{Brightness temperatures and Doppler boosting}

Our VLBI observations reveal a distinct contrast between CSOs and other radio-loud AGN classes in their relativistic beaming properties. 
CSO components typically exhibit moderate brightness temperatures ($T_\mathrm{b} \sim 10^7 - 10^{10}$ K) well below the inverse Compton limit of $\sim 10^{12}$ K \citep{1969ApJ...155L..71K, 1994ApJ...426...51R}. 
This, combined with the derived Doppler-boosting factor  of$<1$, distinguishes CSOs from highly beamed sources such as blazars \citep{2005AJ....130.2473K, 2005AJ....130.1418J, 2006ApJ...642L.115H, 2009A&A...494..527H, 2018ApJ...866..137L,2020ApJS..247...57C}. 
Unlike flat-spectrum radio quasars with a core $T_\mathrm{b} > 10^{11}$ K and Doppler factor of $>10$ \citep{2012A&A...544A..34P}, our CSO sample consistently falls below these thresholds \citep{1996ApJ...460..612R, 2003PASA...20...69P, 2012ApJ...760...77A}, reinforcing their interpretation as young, intrinsically symmetric sources. Because CSOs have moderate brightness temperatures and little Doppler boosting, the observed $T_{\rm b}$ closely tracks intrinsic emissivity, enabling beaming-independent estimates of jet energy density, size, and age.

\subsubsection{Variability}
\label{sec4.5}

Flux density variability in CSOs is generally low \citep{2001AJ....122.1661F, 2024Galax..12...25S} compared to blazars and other core-dominated AGN, which often vary by factors of 2--20 due to relativistic beaming effects \citep[e.g., III Zw 2,][]{2023ApJ...944..187W}. This stability reflects the dominance of lobe emission \citep{1998PASP..110..493O, 2016AN....337....9O} and is a key discriminator from highly variable flat-spectrum sources. 
For the confirmed CSOs with available multi-epoch observations, their total flux densities across the epochs are presented in Fig. S5.
In this work, we quantified whole-source variability as 
$ V = (S_{\rm max} - S_{\rm min})/(S_{\rm max} + S_{\rm min}) $ using flux densities from epochs with matched ($u, v$) ranges where feasible. 
S-band data were prioritised to mitigate short ($u, v$) spacing losses. 
Typical calibration systematics are folded into the $V$ uncertainty via quadratic error propagation.
Across the confirmed CSOs, $V$ is generally $<20\%$ (Table~\ref{tab:cso_diag_summary}), which is consistent with lobe-dominated emission.
Individual components can fluctuate more strongly (Table S7, Fig. S5) but  do not dominate the integrated flux densities. Apparent outliers are often attributable to mismatch in short ($u, v$) spacing or data-quality issues and are excluded from the quantitative $V$ estimate.

The observed size changes in some components correlate with flux-density variations in a pattern consistent with adiabatic expansion of hotspots, where increasing size corresponds to decreasing surface brightness \citep[e.g.,][]{2013A&A...550A.113W}. To enhance measurement reliability, we implemented a consistent model-fitting strategy across epochs, using fixed component numbers where appropriate and excluding measurements from epochs with inadequate data quality. The overall stability of CSO emission aligns with theoretical expectations, as their radio emission originates predominantly from extended lobes and hotspots rather than from variable cores, and relativistic beaming effects are minimal due to the large viewing angles typical for these sources.

\begin{figure*}
\centering
    \includegraphics[width=0.39\linewidth]{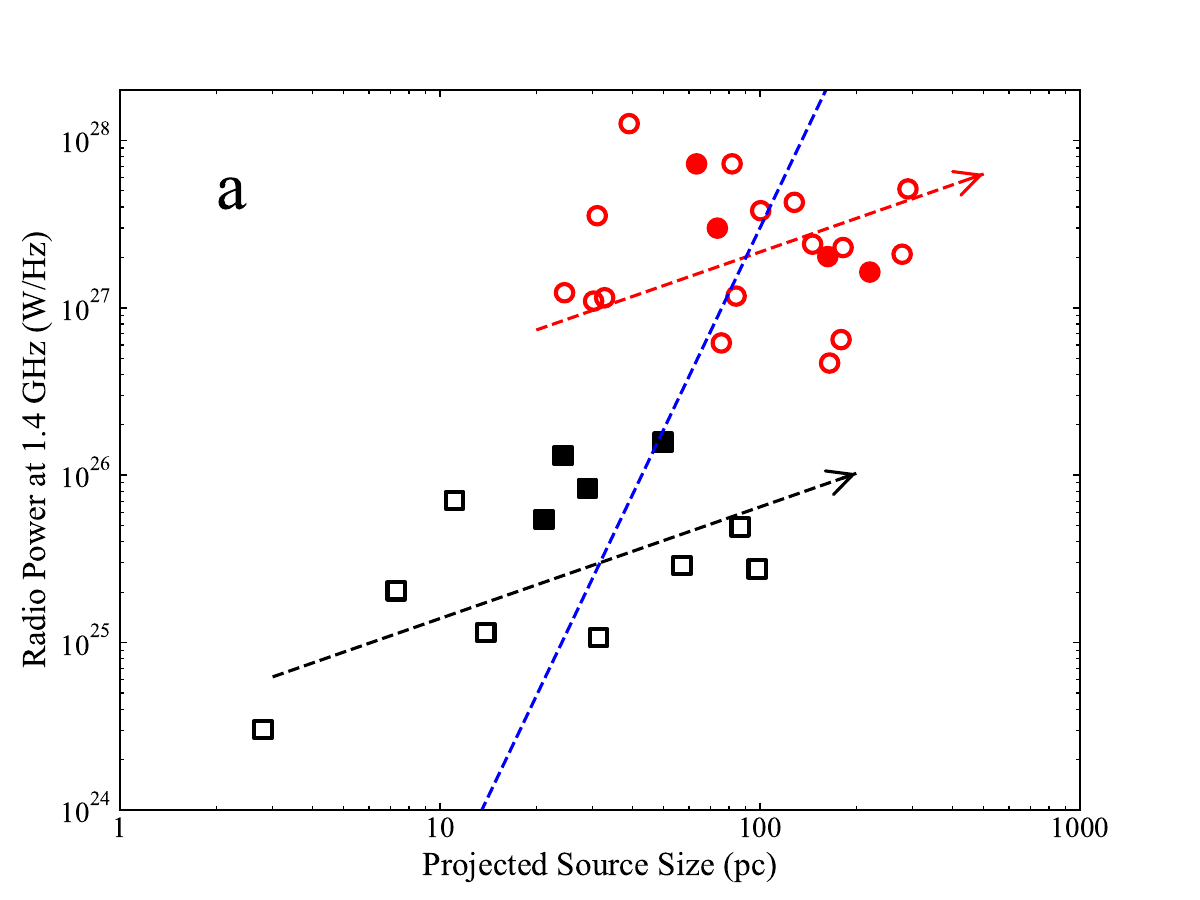}
    \includegraphics[width=0.39\linewidth]{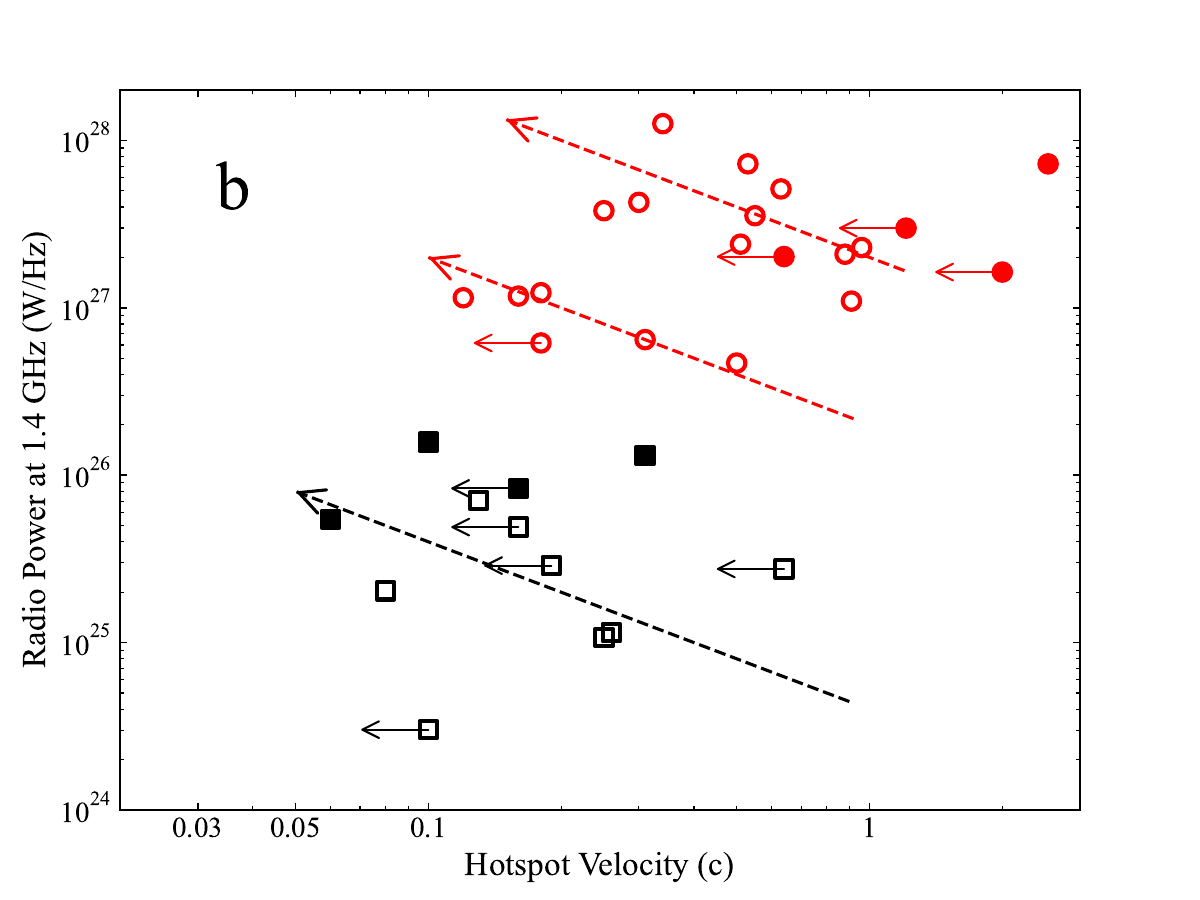} \\
    \includegraphics[width=0.39\linewidth]{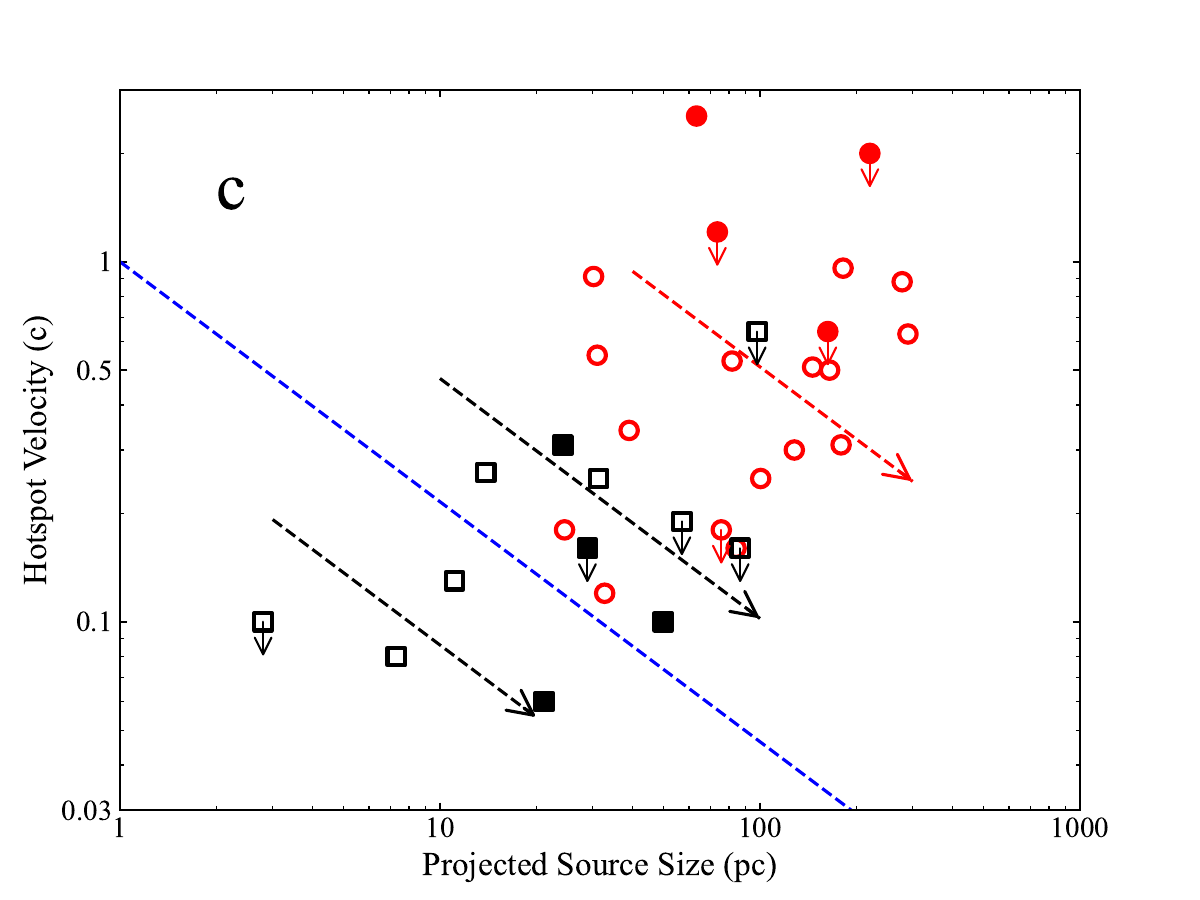}
    \includegraphics[width=0.39\linewidth]{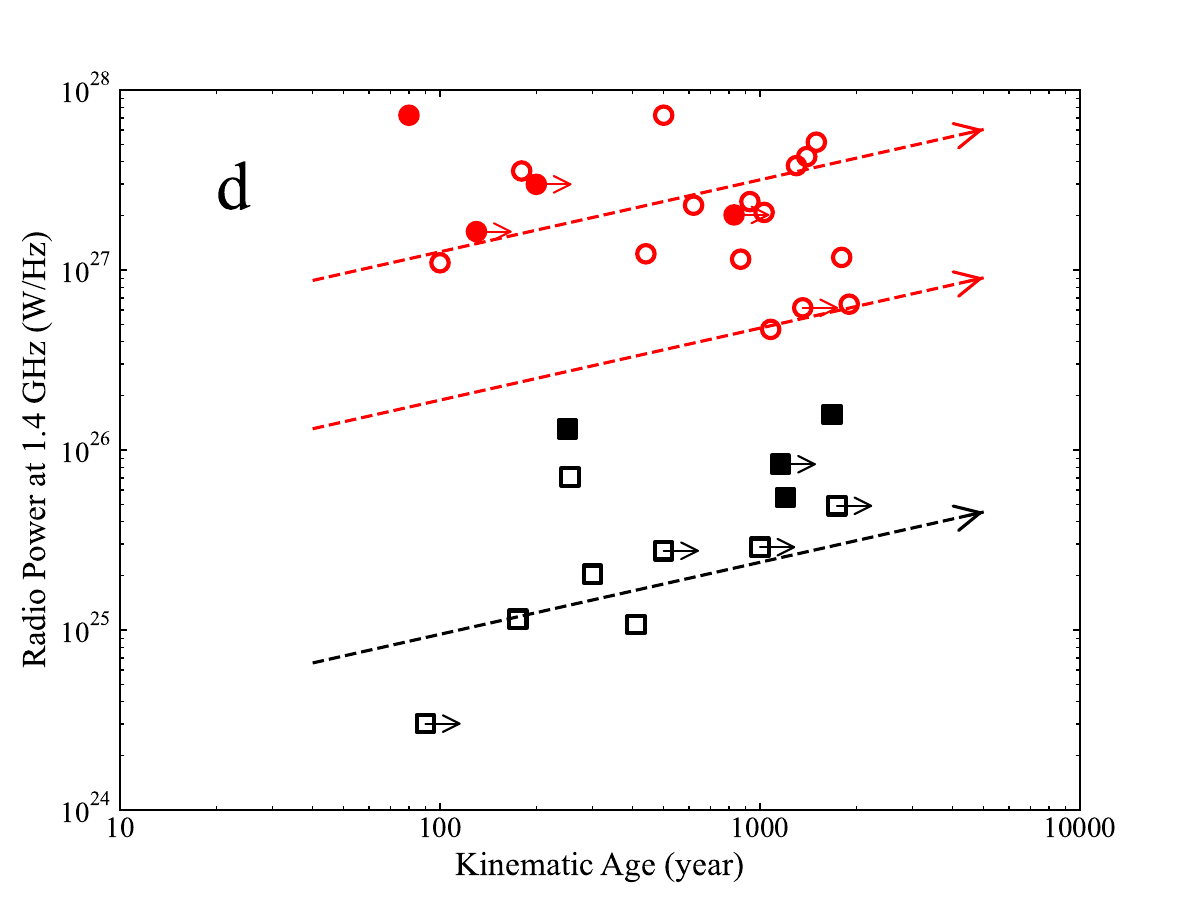} \\
    \includegraphics[width=0.39\linewidth]{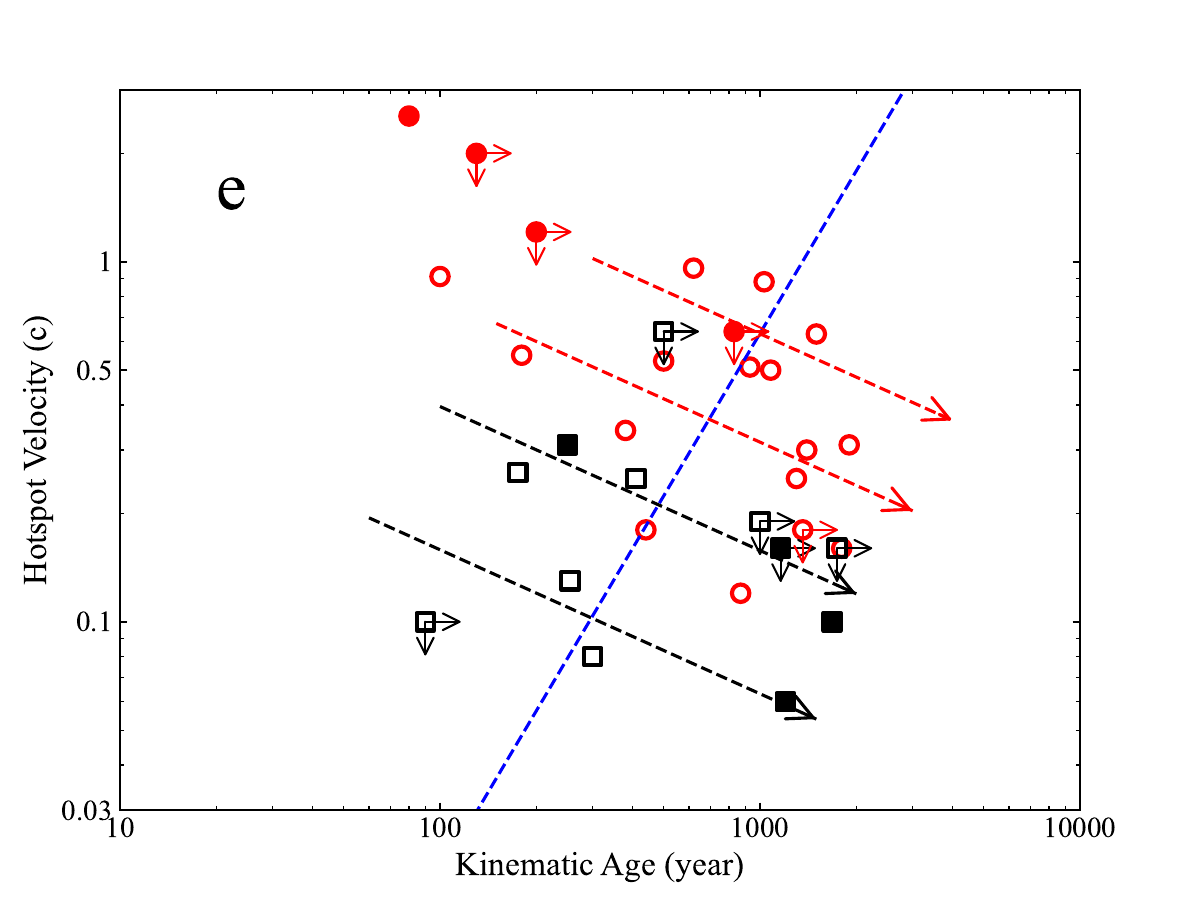}
    \includegraphics[width=0.39\linewidth]{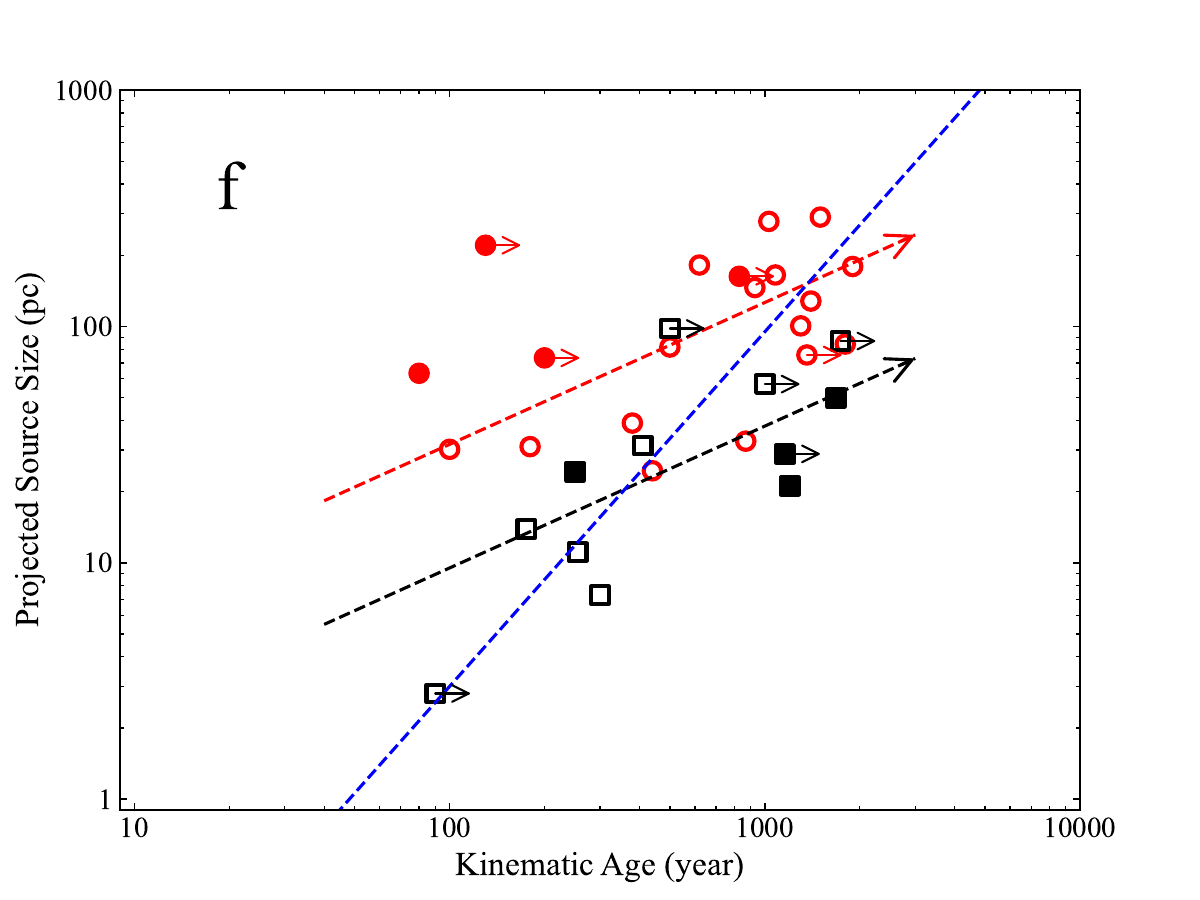} \\
    \includegraphics[width=0.39\linewidth]{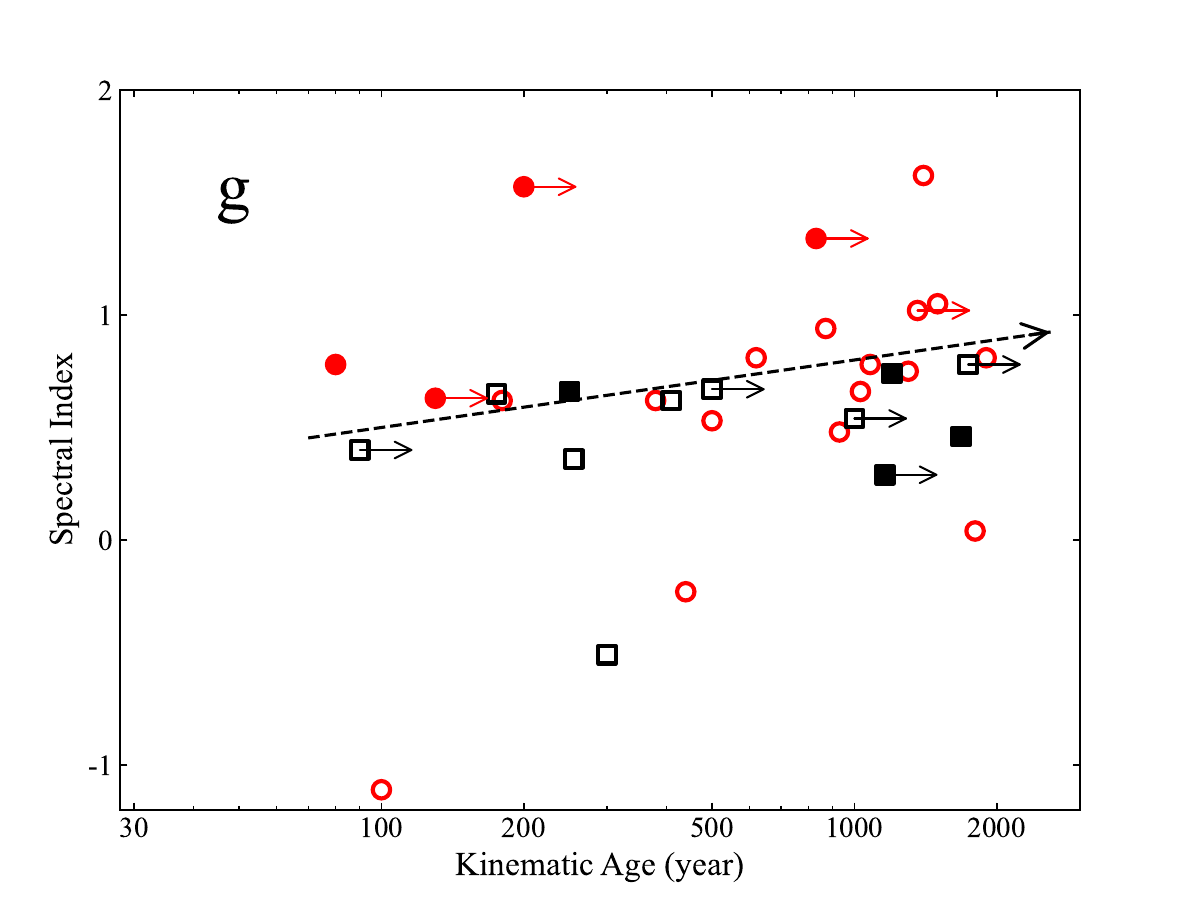}
    \includegraphics[width=0.39\linewidth]{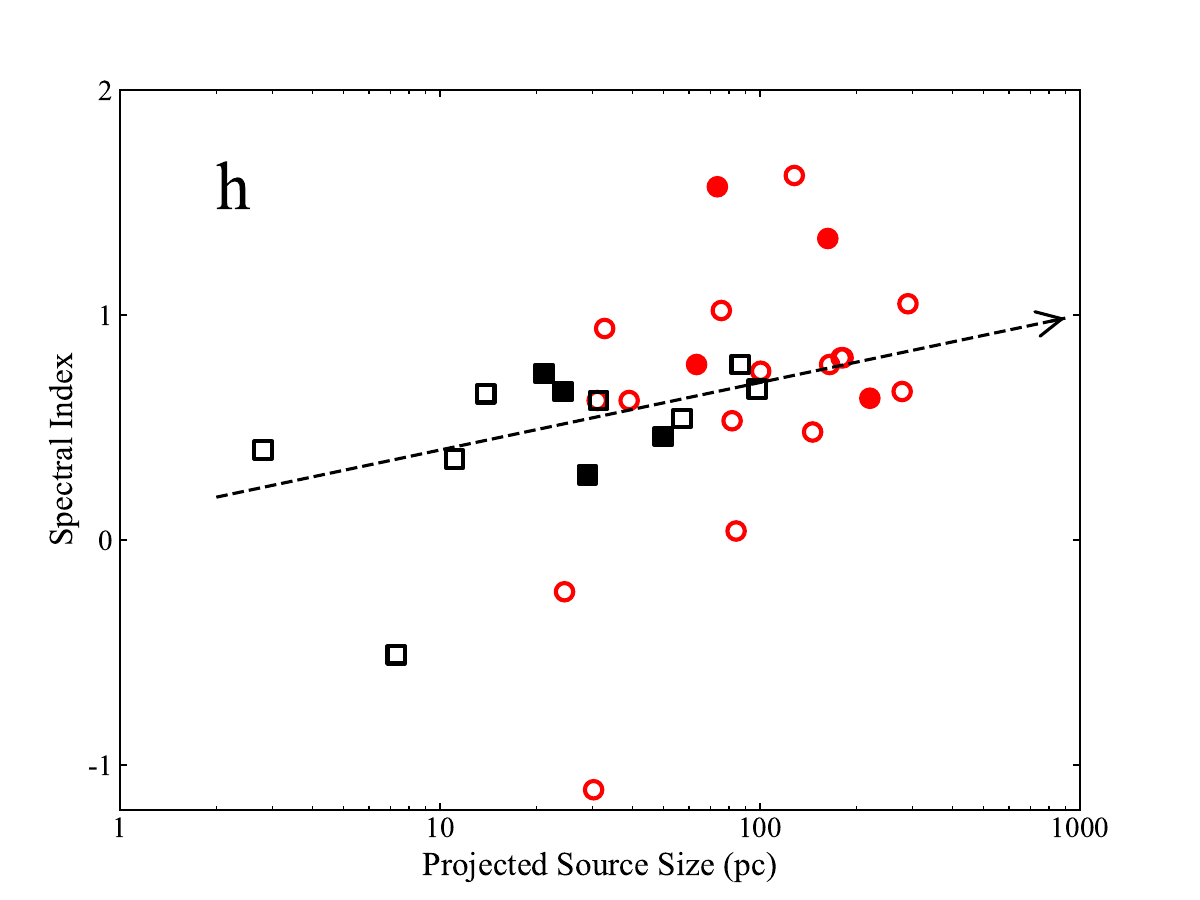}
\caption{Parameter distributions of sample of 32 CSO sources combined with model predictions. 
Black symbols denote low-power CSOs ($P_{\rm rad}<10^{26.5}$ W~Hz$^{-1}$) and red symbols represent high-power CSOs. The filled symbols are from the current study and the open symbols are from \citet{2012ApJ...760...77A}.  
Dashed red and black arrows relate to evolutionary trend for high- and low-power source distributions.
The dashed blue lines in panels {\bf a}, {\bf c}, and {\bf f} show the general power trend of the groups as a whole. 
The dashed blue line in panel {\bf b} shows the expected survival threshold for low power sources.
Panel $\bf a$: Dashed red and black arrows denote the predicted $P_{\rm rad} \propto D^{2/3}$ adiabatic-loss-dominated evolution of a source with constant power, and the dashed blue line indicates the predicted power-size distribution: $P_{\rm rad} \propto D^4$. 
Panel $\bf b$: Dashed red and black arrows show the predicted inverse relationship between radio power and hotspot velocity: $P_{\rm rad} \propto V_{\rm HS}^{-1}$. 
Panel $\bf c$: Hotspot velocity varies with the projected source size as $V_{\rm HS} \propto D^{-2/3}$.
Panel $\bf d$: Dashed arrows show the predicted evolutionary relation of $P_{\rm rad} \propto T^{2/5}$. 
Panels $\bf e$ and $\bf f$: Dashed arrows indicate the predicted evolution following $V_{\rm HS} \propto T^{-2/5}$ and $D \propto T^{3/5}$. 
Panels g and h: Relatively flat distribution of spectral index shows a slight steepening toward $\alpha = 1$ with increasing kinematic age, $T,$ and with projected source size, $D$.
    \label{ab2012_relations}
    }
\end{figure*}

\subsubsection{Hotspot advance speeds and kinematical ages}

\begin{table*}[h]
    \centering
    \caption{Proper motions in CSOs and some CSOcs.}
    \begin{tabular}{cccccc}  \hline \hline
Name            & Comp.  &  $\mu$ &$\beta_{\rm app}$& PA$_{\rm var}$  & Age  \\
                &     & (mas yr$^{-1}$) &(c)& ($\degr$ yr$^{-1}$) & (yr) \\
\hline 
\multicolumn{6}{c}{CSO} \\
\hline
J0832+1832 & W-E & 0.006$\pm$0.004 & 0.06$\pm$0.04 & $-0.01\pm$0.06 & 1200$\pm$820 \\
J1110+4817 
 & SW-NE1 & $-0.078\pm$0.015 & $-3.22\pm$0.64 & $-0.02\pm$0.08 & $>$830 \\
J1111+1955 & W-E & 0.295$\pm$0.062 & 5.55$\pm$1.16 & $-0.02\pm$0.31 &  \\
J1158+2450 & NE-C & 0.001$\pm$0.006 & 0.01$\pm$0.08 & $-0.53\pm$0.07 & $>$1160 \\
 & NW-C & 0.032$\pm$0.007 & 0.42$\pm$0.09 & 0.06$\pm$0.21 & 80$\pm$20 \\
J1234+4753 
 & SE-C & 0.002$\pm$0.003 & 0.06$\pm$0.08 & $-0.01\pm$0.06 & $>$1840 \\
 & SE-NW & 0.004$\pm$0.004 & 0.10$\pm$0.08 & $-0.02\pm$0.05 & 1680$\pm$1400 \\
J1244+4048 & SW2-C & 0.025$\pm$0.045 & 1.10$\pm$2.00 & 0.15$\pm$0.25 & $>$130 \\
J1247+6723 & SE-NW & 0.046$\pm$0.010 & 0.33$\pm$0.07 & 0.03$\pm$0.15 & 160$\pm$30 \\
J1358+4737 & W-E & 0.021$\pm$0.004 & 0.31$\pm$0.06 & 0.07$\pm$0.10 & 250$\pm$50 \\
J1407+2827 & W-E & 0.042$\pm$0.003 & 0.22$\pm$0.02 & $-0.24\pm$0.04 & 160$\pm$10 \\
J1511+0518 & W-E & 0.010$\pm$0.002 & 0.06$\pm$0.01 & 0.04$\pm$0.09 & 480$\pm$100 \\
J1602+2418 & W-E & 0.033$\pm$0.006 & 2.54$\pm$0.45 & 0.43$\pm$0.12 & 80$\pm$10 \\
J1815+6127 
 & SW-NE1 & $-0.082\pm$0.033 & $-2.88\pm$1.16 & $-0.78\pm$0.44 & $>$160 \\
 & NE2-SW & $-0.073\pm$0.035 & $-2.56\pm$1.21 & $-0.27\pm$0.35 & $>$200 \\
J1945+7055 & SW2-NE3 & 0.021$\pm$0.006 & 0.14$\pm$0.04 & $-0.01\pm$0.02 & 1030$\pm$270 \\
 & SW1-NE2 & 0.013$\pm$0.004 & 0.08$\pm$0.03 & $-0.13\pm$0.06 & 720$\pm$220 \\
 & SW2-NE2 & 0.058$\pm$0.006 & 0.38$\pm$0.04 & $-0.08\pm$0.03 & 280$\pm$30 \\
\hline
\hline
    \end{tabular} \\
    \label{tab:pm_fit}
    \tablefoot{
    Column (1): source name;
    Column (2): selected component with respect to the reference component used for proper-motion measurement;
    Column (3): radial proper motion; positive values refer to the components moving away from each other;
    Column (4): apparent radial separation speed measured in the units of $c$ in the source rest frame;
    Column (5): position-angle change, measured from north to east;
    Column (6): kinematical age of the component. If a component has negative or negligible proper motion, we used 1-$\sigma$ of the separation velocity to derive a lower limit \citep[see][]{2012ApJS..198....5A}.
    The fit proper-motion plots are shown in Fig. S6.    \\
    }
\end{table*}

Based on two decades of VLBI monitoring, we measured hotspot separation speeds of $0.06\,c$--$2.54\,c$ (sub-relativistic), consistent with young, expanding sources \citep{2006MNRAS.370.1513K, 2018MNRAS.475.3493B}.
We excluded J1111+1955 owing to inconsistent apparent speeds: $\sim5.5\,c$ (2015--2018; our measurement) versus $<0.19\,c$ (1997--2002) \citep{2005ApJ...622..136G}. Given its Seyfert~II host at $z=0.299$ \citep{1995PASP..107..803U}, where large viewing angles disfavour superluminal patterns, and the sparse cadence, we defer interpretation pending uniform, multi-epoch monitoring.
Two $z>1.5$ quasars (J0943+1702, J1602+2418) show $V_{\rm HS}>2\,c$ separation rates with large uncertainties, plausibly reflecting mildly relativistic jets seen at modest viewing angles. Flux-density-limited selection at high-redshift likely favours such moderately beamed systems.
J1110+4817 and J1815+6127 exhibit apparent ``contraction'' (negative speeds) with bent jet heads, which is consistent with projection and/or pattern-speed effects in curved jets \citep{2012ApJS..198....5A}.
For sources with detected cores, near--far side asymmetries in hotspot speeds likely point to unequal ambient densities on opposite sides of the nucleus \citep{2005A&A...432..823J,2014MNRAS.438..463O}, providing a direct probe of host-galaxy properties in shaping CSO evolution. These diversities stress the need for long-term, same-frequency VLBI to mitigate component misidentification and core-shift systematics \citep{1995MNRAS.277..331S,1998A&A...337...69O}.

Derived kinematic ages span $\sim80$ to $\gtrsim1800$~yr (Table~\ref{tab:pm_fit}); the apparently youngest objects include the two high-$z$ quasars, consistent with flux-limited biases discussed above. The wide age range indicates that we are observing multiple early evolutionary stages \citep{2003PASA...20...69P,2005ApJ...622..136G,2012ApJ...760...77A}. The data favour a continuum from genuinely young, growing systems to sources slowed or stalled by dense environments.

Kinematic ages (see Table ~\ref{tab:pm_fit}) are considered indicators of the order of magnitude. Key caveats are (i) implicit homogeneity and constant-speed assumptions; (ii) asymmetric and clumpy ISM, inducing side-to-side and temporal speed changes; (iii) opacity and core-shift effects when motions are referenced to the core; and (iv) uncertainties in component identification. Despite these limitations, the ages correlate qualitatively with source size (see Section \ref{sec:nature-evolution}), supporting the youth scenario while highlighting environmental regulation of growth.

\subsection{Nature and evolution of CSOs}\label{sec:nature-evolution}

\subsubsection{Evolutionary tracks and classification}

Our kinematic analysis combines 24 CSOs from \citet{2012ApJ...760...77A} with 8 additional sources from this paper, all having measured redshifts and hotspot speeds (Table \ref{tab:ab2012fig}). Together these 32 objects define the working sample. The relationships in Fig.~\ref{ab2012_relations} offer insights into connections between radio power, size, velocity, and age in young radio galaxies.

The power-size correlation (Fig.~\ref{ab2012_relations}-a) demonstrates clear dichotomy in the CSO population, supporting our two-track evolutionary model \citep{2012ApJ...760...77A}. 
A practical threshold at $P_{\rm rad} = 10^{26.5}$~W\,Hz$^{-1}$ \citep{2010MNRAS.408.2261K} separates two regimes: above this power, CSOs show a tight correlation with larger sizes, suggesting efficient conversion of jet power into source expansion; below, sources exhibit a more scattered distribution with limited size growth, indicating environmental influence.
This dichotomy mirrors the FR I/II division \citep{1974MNRAS.167P..31F, 2020NewAR..8801539H}, reflecting models where high-power jets penetrate ambient medium, while low-power jets face environmental impedance.
Although proper-motion requirements are biased in favour of brighter systems, the separation persists across parameters, arguing against a selection artifact.

Self-similar models for young sources expanding into roughly uniform media predict $P_{\rm rad}\propto D^{2/3}$ under nearly constant jet power \citep[e.g.,][]{2007MNRAS.381.1548K}; we therefore used $P$--$D$ scalings as diagnostics rather than strict self-similar laws.  Such models rely on simplifying assumptions including geometric self-similarity, straightforward pressure balance, and quasi-adiabatic magnetic evolution, which may not hold in realistic environments \citep[e.g.,][]{2024ApJ...961..242R}. We therefore treated $P_{\rm rad} - D$ scaling as diagnostic rather than strict laws. An apparently steeper trend in a subset (sometimes described as $P_{\rm rad} \propto D^4$) is empirical and likely shaped by population mixing, environmental diversity, and selection effects. In particular, external pressure gradients can regulate jet collimation and advance speeds \citep[e.g.,][]{2025ApJ...980..119B}, potentially dominating over internal jet microphysics. Jets may be collimated by accretion-disk winds, even on sub-parsec scales, which challenges the concept of strict self-similar evolution. From this perspective, the observed $P-D$ trends reflect how jets interact with ambient pressure profiles rather than idealised self-similar growth.

Figure~\ref{ab2012_relations} panels $b$--$f$ show systematic differences between high-power (red symbols) and low-power (black symbols) CSOs. High-power sources display larger sizes, higher hotspot speeds, and broader age distributions. Radio powers follow theoretical predictions $P_{\rm rad} \propto D^{2/3}$ and $P_{\rm rad} \propto T^{2/5}$ (where $D$ is projected size and $T$ is age), which is consistent with increasing dynamical work done against the ambient medium as adiabatic losses come to dominate.
High-power CSOs ($P > 10^{26.5}$~W\,Hz$^{-1}$) achieve faster hotspot speeds ($> 0.2\,c$) that increase with source size, characterizing young galaxies with sufficient energy flux to penetrate host environments \citep{2006MNRAS.370.1513K}. Low-power CSOs show consistently slower speeds ($< 0.1\,c$) regardless of size, suggesting ISM-dominated expansion \citep{2021A&ARv..29....3O}.
Spectral evolution (Fig.~\ref{ab2012_relations} panels $g$ and $h$) follows the expected patterns: smaller/youngr CSOs exhibit flatter spectra ($\alpha \approx 0.3-0.5$) due to synchrotron self-absorption, while larger/older sources steepen toward $\alpha \approx 1.0$ as absorption weakens with expansion and radiative aging accumulates.

We examined relationships among key CSO parameters using Pearson correlation analyses (see Sect. S7). 
Radio power correlates moderately with both projected size and hotspot velocity ($r = 0.48$), implying that more powerful jets more effectively drive expansion through the ambient medium, producing larger sources and faster hotspots. Hotspot velocity anti-correlates with kinematic age ($r = -0.44$), pointing to systematic deceleration as CSOs grow. 
Projected size correlates positively with spectral index ($r = 0.38$), consistent with synchrotron aging: larger sources host older electrons that have undergone greater radiative and adiabatic losses, leading to steeper spectra.

\subsubsection{Evolutionary pathways and feedback mechanisms}

The CSO population appears to follow power-dependent evolutionary paths. High-power CSOs sustain growth as jets penetrate dense media, evolving along the CSO~$\rightarrow$~MSO~$\rightarrow$~LSO sequence (MSO/LSO: medium/large-sized symmetric objects). Their broader age distribution implies longer survival with double-lobe morphology, and they provide mechanical feedback that evacuates nuclear gas and can influence host evolution on larger scales \citep{2016A&ARv..24...10T,2021A&ARv..29....3O}.

By contrast, low-power CSOs are typically smaller and expand more slowly, likely environment-limited \citep{2025ApJ...980..119B}, often stalling within the ISM. Their narrower size but broader age distributions suggest intermittent or frustrated growth, yielding primarily localized thermal/ISM regulation. Recent work on low-power radio galaxies indicates power-dependent feedback channels with implications for galaxy evolution \citep{2023A&ARv..31....3B,2024MNRAS.529.1472C}.

The sizable number of lower power CSOs supports a frustrated/confined scenario in which weak jets repeatedly interact with a clumpy ISM, injecting energy in nuclear regions yet failing to develop large-scale radio structures. This successful versus frustrated dichotomy aligns with observations of faint sources embedded in tori or circumnuclear star-forming regions, implying local rather than halo/IGM-scale feedback \citep{2020NewAR..8801539H}.

Low-power CSOs also show entrainment and energy losses that erode clear lobes \citep{1982A&A...113..285N,2008ApJ...687..141K,2009MNRAS.397.1113W}, potentially linking to radio-weak AGN \citep{2002MNRAS.329..227S,2019A&A...622A..10C,2024MNRAS.534.1107D}. They are most readily observed in nearby hosts (e.g., J0048+3157/NGC 262; J2355+4950), and some radio-quiet quasars show weak, CSO-like morphologies \citep{2022ApJ...936...73A,2023MNRAS.525..164C,2023MNRAS.525.6064W}. A fraction of high-power CSOs may still deviate from their evolutionary paths, becoming dying CSOs/MSOs due to nuclear activity changes or environment interactions \citep{2010MNRAS.408.2261K,2012ApJ...760...77A}, and may overlap phenomenologically with CSO class 2.2 \citep{2024ApJ...961..242R}.

\subsection{Classification and origin mechanisms}

Recent work proposes a power-based CSO dichotomy similar to ours, with edge-dimmed, low-luminosity (CSO~1) and edge-brightened, high-luminosity (CSO~2) classes \citep{2024ApJ...961..240K}. We adopted the low-power/weak-jet class in line with \citet{2024ApJ...961..242R}; such systems likely dominate the radio-galaxy population and include many newly recognised young sources from low-frequency surveys \citep[e.g.,][]{2020NewAR..8801539H, 2021A&ARv..29....3O}.

\citet{2024ApJ...961..242R} argued that many high-power CSOs are transient, powered by TDEs of giant stars \citep{1988Natur.333..523R}. Our analysis does not support TDEs as the primary engine for most high-power CSOs on several grounds. Concerning 
(1) \emph{hosts}, TDEs prefer spiral galaxies \citep{2017ApJ...850...22L,2020SSRv..216...32F}, whereas CSOs typically inhabit luminous elliptic or ultraluminous infrared galaxies with classical bulges and higher BH masses, often in post-merger environments \citep{2010ApJ...713.1393W,2011MNRAS.412..960T,2020ApJ...897..164K,2020MNRAS.491...92L}. With regard to 
(2) \emph{morphology and continuity}, VLBI frequently shows smoothly connected inner jets feeding edge-brightened hotspots, with inner-jet speeds exceeding hotspot speeds \citep{2000ApJ...541..112T,2012ApJS..198....5A}; this implies sustained power, not rapidly fading transients. 
In terms of (3) \emph{timescales}, CSO lifetimes of $\sim$20--2000~yr (this work) far exceed typical TDE accretion phases (months--years) \citep{1988Natur.333..523R,2021ARA&A..59...21G}; even longer-lived TDE disks ($\sim$10~yr) \citep{2016MNRAS.461..948M} remain orders of magnitude too short. 
Concerning (4) \emph{energetics}, sustained jet powers of $10^{43}$--$10^{47}$~erg\,s$^{-1}$ over centuries to millennia (total up to $\sim$7~M$_\odot c^2$; \citealt{2024ApJ...961..242R}) greatly exceed typical TDE yields ($\sim$0.1~M$_\odot c^2$; \citealt{2016MNRAS.455..859S}). Multiple massive TDEs in quick succession lack support. 
Regarding (5) \emph{rates}, the CSO birth rate ($\sim$3$\times$10$^{-5}$~Gpc$^{-3}$\,yr$^{-1}$; \citealt{1996ApJ...460..612R,2024ApJ...961..242R}) is $\gtrsim$2 orders of magnitude below the TDE rate ($\sim$10$^{-5}$~Mpc$^{-3}$\,yr$^{-1}$), and even if only $\sim$10\% produce jets \citep{2020SSRv..216...81A}, the mismatch persists.

Our interpretation of high-power CSO evolution also differs; whereas \citet{2024ApJ...961..241K} posit that most CSO~2s remain compact ($D \leq 500$~pc, $T \leq 5000$~yr), our data indicate many are \emph{\emph{early stage FR~II}} systems, with some potentially growing to kpc--Mpc scales; several sources with $T>10^3$~yr and $D>500$~pc show continued edge-brightened growth despite moderate hotspot speeds \citep{2003PASA...20...69P,2012ApJ...760...77A}. Additional AGN-like strong optical lines \citep{1994ApJS...91..491G}, infrared excess from dust heating \citep{2011MNRAS.412..960T}, and obscured-AGN X-ray spectra \citep{2006A&A...446...87G,2016ApJ...823...57S,2019ApJ...871...71S} further support long-lived accretion. Extended radio emission beyond CSO scales in some objects (e.g., 0108+388, 1511+0518) argues for recurrent AGN activity \citep{1990A&A...232...19B,2008A&A...487..885O,2010MNRAS.408.2261K}.

A more plausible engine is standard AGN accretion, fuelled by merger-driven inflows or bar instabilities \citep{2010MNRAS.407.1529H,2019NatAs...3...48S}, which is consistent with the prevalence of interacting hosts and the high fraction among IR-luminous galaxies. While rare CSOs might arise from exceptional TDEs, population-level properties favour conventional AGN. Selection biases from \textit{Gaia} (optically brighter, lower-$z$ systems) are acknowledged, but radio evolutionary trends appear consistent across \textit{Gaia}-detected and non-detected subsamples.

Forthcoming Square Kilometre Array (SKA) surveys will uncover weaker CSOs and clarify the transition around $P_{\rm 1.4\,GHz} \lesssim 10^{25}$~W\,Hz$^{-1}$. SKA pathfinders already improve demographic baselines for FR~I/II samples \citep[e.g.,][]{2019A&A...622A..12H,2019MNRAS.488.2701M,2020PASA...37...18W,2020PASA...37...17W,2023MNRAS.518.4290S}. Ultimately, capturing the \emph{\emph{full}} CSO population, rather than only bright or peculiar subsets, will reveal more diverse evolutionary paths and sharpen constraints on radio-galaxy growth and AGN feedback.

\section{Summary} \label{sec:summary}

In this study, we developed and validated a \textit{Gaia}+VLBI method to identify and characterize CSOs. Our main findings are listed as follows.

\begin{enumerate}
    \item The \textit{Gaia}+VLBI approach decisively classifies 30 out of 40 candidates: 20 confirmed CSOs (15 via direct \textit{Gaia}--VLBI alignment; 5 via morphology/spectra despite larger offsets) and ten core--jet sources; ten remain ambiguous. This overcomes the long-standing difficulty of locating faint/absorbed radio cores in young sources. Scaling this \textit{Gaia}+VLBI framework to larger samples, especially with sensitive SKA-era observations of low-power CSOs, will reveal hidden populations and sharpen constraints on feedback, frustrated sources, and host-galaxy-jet coupling across cosmic time.
    
    \item The two-track evolutionary paradigm is reinforced by multiple, independent observational correlations presented here. High-power CSOs (\(P_{\rm rad}>10^{26.5}\,{\rm W\,Hz^{-1}}\)) show moderate hotspot speeds (typically \(0.2 - 0.5\,c\)), systematic spectral gradients, and potential growth into large radio galaxies. Low-power CSOs remain sub-kiloparsec with slower expansion (\(<0.1\,c\)) and strong environmental imprint, consistent with the \emph{\emph{frustrated radio galaxy}} picture. Correlations among radio power, size, hotspot speed, and kinematic age match theoretical expectations, implying distinct trajectories set largely by initial radio power.
    
    \item The properties of CSOs suggest that isolated TDEs are unlikely to be the dominant origin: long activity, persistent symmetry, and energy budgets in some CSOs instead point to standard AGN fueling.
    
    \item Beyond power, evolution reflects environment and duty cycle: ambient density and magneto-ionic conditions, intermittent accretion, engine lifetime shape morphology, spectra, and growth, hence the observed diversity of observational characteristics.
    
    \item Compact symmetric objects are laboratories for AGN feedback: high-power systems deliver mechanical work on larger scales; low-power systems heat and disturb the central ISM without developing large lobes.
    
\end{enumerate}

\section{Data availability}

The VLBI datasets underlying this article were derived from the public domain in the Astrogeo archive (\url{http://astrogeo.org/}). 
The materials originally presented as appendices are available as online supplementary material at Zenodo (DOI: \url{https://doi.org/10.5281/zenodo.17278723}). Figure and table numbering in the supplement follows the S-scheme (e.g. Fig. S1, Table S1).

  \begin{acknowledgements}
We thank the referee for the constructive comments that improved the paper. TA and WAB are grateful to Anthony C. Readhead for his constructive discussions and valuable suggestions.
This work is supported by the National Key R\&D Program of China (2022SKA0120102).
This work used resources of China SKA Regional Centre prototype funded by the Ministry of Science and Technology of the People's Republic of China and the Chinese Academy of Sciences \citep{2019NatAs...3.1030A,2022SCPMA..6529501A}.
TA and WAB acknowledge the support from the Xinjiang Tianchi Talent Program.
YKZ is supported by the China Scholarship Council (No. 202104910165),
the Shanghai Sailing Program under grant number 22YF1456100, and the Strategic Priority Research Program of the Chinese Academy of Sciences (Grant No. XDA0350205).
YKZ thanks for the warm hospitality and the helpful comments from Ivy Wong in CSIRO Space\&Astronomy in Australia. 
YKZ thanks for the valuable suggestions and comments from Krisztina \'Eva Gab\'anyi in Konkoly Observatory, Hungary.
SF thanks the Hungarian National Research, Development and Innovation Office (NKFIH, grant no. OTKA K134213) for support. 
This work was also supported by the NKFIH excellence grant TKP2021-NKTA-64.
The authors acknowledge the use of Astrogeo Center database maintained by L. Petrov.
This work has made use of data from the European Space Agency (ESA) mission \textit{Gaia} (\url{https://www.cosmos.esa.int/gaia}), processed by the \textit{Gaia} Data Processing and Analysis Consortium (DPAC, \url{https://www.cosmos.esa.int/web/gaia/dpac/consortium}).
This work has made use of the NASA Astrophysics Data System Abstract Service, and the NASA/IPAC Extragalactic Database (NED), which is operated by the Jet Propulsion Laboratory, California Institute of Technology, under contract with the National Aeronautics and Space Administration.
The National Radio Astronomy Observatory is a facility of the National Science Foundation operated under cooperative agreement by Associated Universities, Inc. 

\end{acknowledgements}

\bibliographystyle{aa} 
\bibliography{refer.bib}

\appendix

\section{Supplementary tables}\label{app:tables}

\begin{table*}
    \caption{The CSO candidate sample collected in this paper.}
    \centering
    \begin{tabular}{cccccc} \hline\hline
Source name & Right ascension (h~m~s), declination (\degr~\arcmin~\arcsec)                & $z$     &Type & Ref. & Classification \\ \hline
J0003+2129 & 00:03:19.350009 +21:29:44.50822 & 0.450 & Q & 2 & CJ \\
J0005+0524 & 00:05:20.215504 +05:24:10.80305 & 1.887 & Q & 2 & CJ \\
J0048+3157 & 00:48:47.141485 +31:57:25.08483 & 0.015 & G & 2 & CSO \\
J0119+3210 & 01:19:35.001084 +32:10:50.06103 & 0.0602 & G & 2 & CSOc \\
J0650+6001 & 06:50:31.254327 +60:01:44.55477 & 0.455 & Q & 1,2 & CSOc \\
J0741+2706 & 07:41:25.732847 +27:06:45.39211 & 0.772 & Q & 3 & CSO \\
J0753+4231 & 07:53:03.337437 +42:31:30.76470 & 3.594 & Q & 1,3 & CJ \\
J0831+4608 & 08:31:39.802592 +46:08:00.77140 & 0.131 & G & 3 & CSOc \\
J0832+1832 & 08:32:16.040301 +18:32:12.13265 & 0.154 & G & 3 & CSO \\
J0906+4636 & 09:06:15.539809 +46:36:19.02416 & 0.0847 & G & 3 & CSOc \\
J0943+1702 & 09:43:17.223952 +17:02:18.96252 & 1.600 & Q & 3 & CSOc \\
J1110+4817 & 11:10:36.324124 +48:17:52.44997 & 0.742 & Q & 3 & CSO \\
J1111+1955 & 11:11:20.065601 +19:55:36.00040 & 0.299 & G & 1,2,3 & CSO \\
J1148+5924 & 11:48:50.358181 +59:24:56.38223 & 0.0108 & G & 1,3 & CSOc \\
J1148+5254 & 11:48:56.569098 +52:54:25.32250 & 1.638 & Q & 2 & CJ \\
J1158+2450 & 11:58:25.787561 +24:50:17.96392 & 0.203 & G & 3 & CSO \\
J1234+4753 & 12:34:13.330774 +47:53:51.23687 & 0.373 & Q & 3 & CSO \\
J1244+4048 & 12:44:49.187531 +40:48:06.16239 & 0.814 & Q & 1,3 & CSO \\
J1247+6723 & 12:47:33.329586 +67:23:16.44894 & 0.107 & G & 2 & CSO \\
J1254+1856 & 12:54:33.271499 +18:56:01.90866 & 0.125 & G & 3 & CSOc \\
J1256+5652 & 12:56:14.233979 +56:52:25.23760 & 0.0417 & G & 2 & CSO \\
J1309+4047 & 13:09:41.508928 +40:47:57.23878 & 2.908 & Q & 2 & CSOc \\
J1310+3403 & 13:10:04.433638 +34:03:09.10854 & 0.960 & G & 3 & CSO \\
J1311+1417 & 13:11:07.824225 +14:17:46.64778 & 1.955 & Q & 1 & CJ \\
J1326+3154 & 13:26:16.511702 +31:54:09.52057 & 0.368 & G & 2,3 & CSO \\
J1335+4542 & 13:35:21.962189 +45:42:38.23218 & 2.451 & Q & 2 & CJ \\
J1358+4737 & 13:58:40.666477 +47:37:58.31155 & 0.230 & G & 3 & CSO \\
J1407+2827 & 14:07:00.394417 +28:27:14.69011 & 0.077 & G & 2 & CSO \\
J1511+0518 & 15:11:41.266365 +05:18:09.25931 & 0.084 & G & 2 & CSO \\
J1559+5924 & 15:59:01.701929 +59:24:21.83416 & 0.060 & G & 2,3 & CSOc \\
J1602+2418 & 16:02:13.838513 +24:18:37.79350 & 1.791 & Q & 3 & CSO \\
J1616+0459 & 16:16:37.556823 +04:59:32.73651 & 3.215 & Q & 2 & CJ \\
J1755+6236 & 17:55:48.435228 +62:36:44.12661 & 0.0276 & G & 2 & CSO \\
J1815+6127 & 18:15:36.792244 +61:27:11.64744 & 0.601 & Q & 1 & CSO \\
J1816+3457 & 18:16:23.901115 +34:57:45.74704 & 0.245 & G & 1,2 & CJ \\
J1823+7938 & 18:23:14.108654 +79:38:49.00188 & 0.224 & Q & 1,2 & CSOc \\
J1945+7055 & 19:45:53.519774 +70:55:48.72880 & 0.101 & G & 1,2 & CSO \\
J2022+6136 & 20:22:06.681748 +61:36:58.80472 & 0.227 & G & 1,2 & CJ \\
J2245+0324 & 22:45:28.284742 +03:24:08.86404 & 1.350 & Q & 1 & CJ \\
J2355+4950 & 23:55:09.458159 +49:50:08.33951 & 0.238 & G & 1,2 & CSO \\
\hline
    \end{tabular}\\
    \tablefoot{
    Column (1): source name.
    Column (2): Equatorial coordinates from Astrogeo \citep{2024arXiv241011794P}.
    Column (3): Spectroscopic or photometric redshifts collected from the NASA/IPAC Extragalactic Database (\url{https://ned.ipac.caltech.edu/}) or SIMBAD (\url{https://simbad.u-strasbg.fr/}).
    Column (4) lists the optical identification of the source type from SIMBAD. Q -- quasar; G -- galaxy. 
    Column (5) lists the references to the sources. 
    Column (6) presents the classification of the radio morphology: 
    CSO  -- confirmed CSO with \textit{Gaia}-identified AGN between two radio components; 
    CJ   -- one-sided core--jet with \textit{Gaia}-identified AGN at the end of the radio structure;
    CSO candidate -- sources with large optical--radio offset remain as CSO candidates. 
    References: 
    1 -- \cite{2000ApJ...534...90P}; 
    2 -- \cite{2012ApJ...760...77A}; 
    3 -- \cite{2016MNRAS.459..820T}.
    }
    \label{tab:sample}
\end{table*}

\begin{table*}[h]
    \centering
    \caption{CSO parameters used in Fig.~2 in the main paper.}
    \begin{tabular}{ccccccc}  \hline \hline
Name            & Radio power $P_{1,4}$   &  Projected size & H--H velocity & Kinematic age & Spectral index & Ref \\
                &  log(W~Hz$^{-1}$)   & (pc) &($c$)& (year) & & \\
\hline 
J0038+2302 & 25.03 & 31.3 & 0.25 & 410 & 0.62 & 2 \\
J0111+3906 & 28.10 & 39.0 & 0.34 & 380 & 0.62 & 2 \\
J0119+3210 & 25.44 & 98.0 & $\le$0.64 & $\ge$500 & 0.67 & 2 \\
J0713+4349 & 27.38 & 146.0 & 0.51 & 930 & 0.48 & 2 \\
J0832+1832 & 25.74 & 21.2 & 0.06 & 1200 & 0.74 & 1 \\
J1035+5628 & 27.36 & 181.8 & 0.96 & 620 & 0.81 & 2 \\
J1110+4817 & 27.31 & 162.9 & $\le$0.64 & $\ge$830 & 1.34 & 1 \\
J1111+1955 & 26.79 & 75.7 & $\le$0.18 & $\ge$1360$*$& 1.02 & 1,2 \\
J1158+2450 & 25.92 & 28.9 & $\le$0.16 & $\ge$1160 & 0.29 & 1 \\
J1234+4753 & 26.20 & 49.9 & 0.10 & 1680 & 0.46 & 1 \\
J1244+4048 & 27.21 & 220.4 & $\le$2.00 & $\ge$130 & 0.63 & 1 \\
J1247+6723 & 25.06 & 13.9 & 0.26 & 160 & 0.65 & 1,2 \\
J1324+4048 & 27.06 & 32.7 & 0.12 & 870 & 0.94 & 2 \\
J1326+3154 & 27.32 & 278.1 & 0.88 & 1030 & 0.66 & 1,2 \\
J1335+5844 & 27.07 & 84.2 & 0.16 & 1800 & 0.04 & 2 \\
J1358+4737 & 26.12 & 24.3 & 0.31 & 250 & 0.66 & 1 \\
J1407+2827 & 25.85 & 11.1 & 0.13 & 160 & 0.36 & 1,2 \\
J1414+4554 & 25.69 & 86.7 & $\le$0.16 & $\ge$1740 & 0.78 & 2 \\
J1415+1320 & 27.04 & 30.2 & 0.91 & 100 & $-$1.11 & 2 \\
J1511+0518 & 25.31 & 7.3 & 0.08 & 480 & $-$0.51 & 1,2 \\
J1602+2418 & 27.86 & 63.4 & 2.54 & 80 & 0.78 & 1 \\
J1609+2641 & 27.71 & 290.0 & 0.63 & 1500 & 1.05 & 2 \\
J1723+6500 & 24.48 & 2.8 & $\le$0.10 & $\ge$90 & 0.4 & 2 \\
J1734+0926 & 27.58 & 100.5 & 0.25 & 1300 & 0.75 & 2 \\
J1815+6127 & 27.48 & 73.6 & $\le$1.21 & $\ge$200 & 1.57 & 1 \\
J1845+3541 & 27.55 & 31.0 & 0.55 & 180 & 0.62 & 2 \\
J1939$-$6342 & 27.63 & 128.0 & 0.30 & 1400 & 1.62 & 2 \\
J1944+5448 & 26.67 & 165.0 & 0.50 & 1080 & 0.78 & 2 \\
J1945+7055 & 25.46 & 57.0 & $\le$0.19 & $\ge$1000 & 0.54 & 1,2 \\
J2022+6136 & 27.09 & 24.5 & 0.18 & 440 & $-$0.23 & 2 \\
J2203+1007 & 27.86 & 81.8 & 0.53 & 500 & 0.53 & 2 \\
J2355+4950 & 26.81 & 179.2 & 0.31 & 1900 & 0.81 & 1,2 \\
\hline
    \end{tabular} \\
    \label{tab:ab2012fig}
    \tablefoot{
    Column (1): source name;
    Column (2): radio power at 1.4 GHz;
    Column (3): projected linear size in pc;
    Column (4): hotspot separation velocity in $c$;
    Column (5): kinematic age;
    Column (6): spectral index calculated from 2--8 GHz from this paper or 5--8 GHz from \citet{2012ApJ...760...77A};
    Column (7): references: 1 -- this paper, 2 -- \citet{2012ApJ...760...77A}; Note: Five sources in the \textit{Gaia}+VLBI CSO sample are also included in \citet{2012ApJ...760...77A}, and for these sources, the data from  \citet{2012ApJ...760...77A} is used.
    $*$: For sources with conflicting kinematic measurements (particularly J1111+1955), we adopt the more reliable values from previous studies with longer monitoring baselines as indicated in the reference column.}
\end{table*}

\end{document}